\documentclass[prl,twocolumn,superscriptaddress]{revtex4-2} 

\usepackage{physics} 
\usepackage{amssymb} 
\usepackage{xcolor} 
\usepackage{graphicx}

\usepackage[T1]{fontenc}
\usepackage{mathptmx} 
\usepackage[cal=cm]{mathalfa}

\definecolor{deepblue}{rgb}{0, 0, 0.7} 
\usepackage[colorlinks]{hyperref}
\hypersetup{
    linkcolor=black,
    citecolor=deepblue, 
}

\newcommand{\sty}{\scriptstyle}
\newcommand{\ssty}{\scriptscriptstyle} 

\newcommand{\move}[1]{\hspace{#1pt}}

\newcommand{\up}[2]{^{\raisebox{#1pt}{$\sty#2$}}} 
\newcommand{\down}[2]{_{\raisebox{-#1pt}{$\sty#2$}}}

\newcommand{\newenv}[2]{\newcounter{#1}
\newenvironment{#1}{\refstepcounter{#1}\par\emph{#2 \arabic{#1}:}}{\par}
\newenvironment{#1*}{\par\emph{#2:}}{\par}
}
\newenv{definition}{Definition}
\newenv{theorem}{Theorem}
\newenv{lemma}{Lemma}
\newenv{corollary}{Corollary}
\newenv{proposition}{Proposition}
\newenv{construction}{Construction}
\newenv{notation}{Notation}

\newcommand{\Newenv}[1]{
\expandafter\newcommand\csname #1s\endcsname[2]{\begin{#1}##1\end{#1}}
\expandafter\newcommand\csname #1ss\endcsname[2]{\begin{#1*}##1\end{#1*}}}
\Newenv{equation} 
\Newenv{definition} 
\Newenv{theorem} 
\Newenv{lemma} 
\Newenv{corollary} 
\Newenv{proposition} 

\newcommand{\NewEnv}[1]{
\expandafter\newcommand\csname #1s\endcsname[2]{\begin{#1}{##1}##2\end{#1}}}
\NewEnv{array} 

\newcommand{\newEnv}[1]{
\expandafter\newcommand\csname #1s\endcsname[2][]{\begin{#1}[##1]##2\end{#1}}}
\newEnv{figure} 

\usepackage{booktabs}
\usepackage{subcaption}
\usepackage{forest}

\usepackage{xstring}
\NewDocumentCommand{\zoom}{ m m }{%
	\IfEqCase{#1}{%
		{c}{\resizebox{\columnwidth}{!}{$#2$}}
		{t}{\resizebox{\textwidth}{!}{$#2$}}
		{p}{\resizebox{\paperwidth}{!}{$#2$}}
		{l}{\resizebox{\linewidth}{!}{$#2$}}
	}[\scalebox{#1}{$#2$}]
}

\DeclareMathOperator{\Span}{span}

\begin{document}
    \title{
    Strong nonlocality with more imaginarity and less entanglement
    }
    
   	\author{Subrata Bera}
	\email{98subratabera@gmail.com}
	\affiliation{Department of Applied Mathematics, University of Calcutta, 92, A.P.C. Road, Kolkata- 700009, India}

	\author{Indranil Biswas}
	\email{indranilbiswas74@gmail.com}
	\affiliation{Department of Applied Mathematics, University of Calcutta, 92, A.P.C. Road, Kolkata- 700009, India}
	
	\author{Atanu Bhunia}
	\email{atanu.bhunia31@gmail.com}
	\affiliation{Physics and Applied Mathematics Unit, Indian Statistical Institute, 203, B.T. Road, Kolkata-700108, India}
	
\author{Indrani Chattopadhyay}
\email{icappmath@caluniv.ac.in}
\affiliation{Department of Applied Mathematics, University of Calcutta, 92, A.P.C. Road, Kolkata- 700009, India}
	
	\author{Debasis Sarkar}
	\email{dsarkar1x@gmail.com,dsappmath@caluniv.ac.in}
	\affiliation{Department of Applied Mathematics, University of Calcutta, 92, A.P.C. Road, Kolkata- 700009, India}
\date{\today}
\begin{abstract}
Complex numbers are central to the formulation of quantum mechanics, yet their role as a genuine resource is only beginning to be understood. In this work, we demonstrate that quantum states with intrinsically complex amplitudes provide a fundamental advantage in state discrimination. We construct a set of five orthogonal three qubit pure states and show that the set is strongly nonlocal if and only if it includes imaginary components. Such a set becomes locally indistinguishable not only under local measurements but also against bipartite joint measurements. This enhanced robustness makes imaginarity a valuable resource for quantum cryptography since information encoded in these states remains secure against collaborative group attacks. Our results highlight a new operational role of complex numbers in quantum theory and establish imaginarity as a key enabler of cryptographic security. However, we reconstruct the set by replacing the only product state with a biseparable state whose shared entanglement between two parties nullifies the effect of imaginarity in exhibiting strong nonlocality. In fact, we show how entangling correlations between two distant parties can dilute the effect of imaginarity, and conversely, how imaginarity itself can mimic the role of entanglement. Nevertheless, the set spans a locally indistinguishable subspace, while its complement, in turn, produces distillable genuine entanglement. Notably, this is the smallest possible Unextendible Biseparable Basis (UBB) that resolves the open problem regarding the existence of a UBB of cardinality $d^2+d-1$ in $d^{\otimes 3}$.  Our construction yields a highly powerful set, rich in resources from multiple perspectives of quantum information theory, including many-copy discrimination, unambiguous identification, entanglement creation from product state, and non-entangling perturbations. 
\end{abstract} 
	
    \maketitle

The use of complex numbers in quantum theory has been a convenient tool since its inception. From Schrödinger’s wave mechanics to the modern Hilbert-space framework, complex amplitudes play a foundational role in this theory. A growing ensemble of research has established that quantum mechanics over real vector spaces is fundamentally weaker than its complex counterpart, with clear differences in tomographic completeness \cite{Hardy2012,Niestegg2020}, entanglement structure \cite{Wootters2012,Yanna2025}, quantum communication \cite{Elliott2025}, quantum steering \cite{Wei2024,Archan2025}, state conversion \cite{Swapan2021PRA} and discrimination \cite{Swapan2021PRL,Zoratti2021}, and channel discrimination \cite{Swapan2024} capabilities. Nevertheless, it has long been debated whether such an imaginary unit reflects a physical necessity or merely a mathematical convenience \cite{Hita2025,Hoffreumon2025}. Recent landmark theoretical and experimental results have clarified that quantum theory restricted to real-valued amplitudes can, in principle, be experimentally falsified \cite{Renou2021,Chen2022}. These findings establish quantum imaginarity as an indispensable component for the precise description and manipulation of quantum systems. In particular, the presence of complex phases can signal nonclassical behavior, in close analogy to entanglement \cite{Horodecki2009,Sarkar2008,Char2021,Biswas2023,Biswas2025,Char2025} and coherence \cite{Baumgratz2014,Winter2016,Streltsov2017RMP,Streltsov2017PRL,Dipayan2019,Char2023,Prabir2024,Char2024}. 

In parallel, the resource theory of imaginarity (RTI) \cite{Hickey2018} has emerged as a systematic framework for capturing the operational significance of complex numbers. In this setting, the theory is specified by a pair $(\mathcal{F},\mathcal{O})$, where the set of free states $\mathcal{F}=\{\rho \mid \rho=\rho^\ast\}$ consists of density matrices with real entries in a fixed basis, and the set of free operations $\mathcal{O}$ includes only real operations that cannot generate imaginarity. Any state with a nonzero imaginary component thus serves as a resource, whose quantification is determined by monotones under $\mathcal{O}$, such as the Robustness of Imaginarity (RoI) \cite{Hickey2018}, Geometric Imaginarity \cite{Swapan2021PRA}, and others \cite{Xu2024,Du2025,Liu2026}. Rather than full state tomography, such imaginarity in certain states can be detected using witness-based \cite{Fernandes2024,Zhang2025} or moment-based \cite{Bivas2026} approaches.

Importantly, the resource character of imaginarity is not merely theoretical but also visible in experimental practice. For example, in polarization-encoded photonic systems, an arbitrary rotation around the $y$-axis can be realized using a single half-wave plate (HWP), whereas a $z$-axis rotation requires the more complex QHQ (quarter-half-quarter wave plate) sequence \cite{Swapan2021PRA}. Such asymmetry in experimental effort highlights that imaginary transformations are intrinsically more costly to implement than their real counterparts, demonstrating the imaginary part of quantum states as an operationally relevant resource rather than only a mathematical abstraction.

Despite significant advances in quantum physics, the precise role of imaginarity as a resource in quantum nonlocality remains largely unexplored. Nonlocality refers to quantum correlations that cannot be reproduced by local hidden variable models \cite{Werner1989,Horodecki1995} and has, over the past decades, become a central concept in characterizing nonclassical phenomena. Beyond the standard Bell-type scenario \cite{Bell1964,Brunner2014,Kundu2020,Kundu2024,Kundu2024EJM,Kundu2025}, nonlocality also arises in the context of state discrimination \cite{Bennett1999PB,Bennett1999UPB,Walgate2000,Walgate2002,Horodecki2003,Sarkar2004, Xin2008,Somshubhro2011,Manik2021,Bhunia2021,Bhunia2022,Bhunia2023,Subrata2024,Bhunia2025,Bhunia2026}. In this framework, a state is chosen from a known set of orthogonal quantum states and distributed among distant parties. The goal is to identify the state using only local operations and classical communications (LOCC). The inability to perform the task is referred to as local indistinguishability. Understanding which structural properties of quantum states give rise to such nonlocal behaviour is therefore of both fundamental interest and practical importance, especially for applications in quantum cryptography and secure communication protocols \cite{Terhal2001,Eggeling2002,Chattopadhyay2007,Markham2008,Matthews2009}. Recently, it has been proven that the use of complex numbers in measurement basis could help to distinguish mixed states. However, for pure states, the use of complex numbers is rather ornamental. With this motivation, we investigate the resourcefulness of imaginarity in pure states. Specifically, we construct a family of orthogonal pure states in $2\bigotimes2\bigotimes2$ system and demonstrate that the presence of imaginary components extends the known boundaries of nonlocality. Remarkably, this set remains secure not only against LOCC-based attacks but also against coordinated group attacks. This set constitutes a well-recognized example of a strongly nonlocal set \cite{Saronath2019,Zhang2019,Yuan2020,Bhunia2024}, for which no state can be eliminated under local as well as joint measurements. This highlights imaginarity as a powerful enabler of cryptographic security. In particular, information encoded in such states resists extraction even with collaborative operations, thereby opening a new avenue for protocols that exploit complex phases as intrinsic shields against eavesdropping. To shed light on the role of imaginarity in state discrimination, step by step construction of the following bases in $2\bigotimes2\bigotimes2$ system acts as a platform to reveal its impact.

\begin{construction}
 Let us define a Cross-L structure as a nonplanar configuration formed by joining two planar L-blocks along a shared edge. The three-qubit tile structure can be decomposed into two such shapes, where each side of an L-block consists of two tiles sharing a common tile. In this setting, each L-block represents two orthogonal biseparable states, and any pair of states drawn from different L-blocks within the same Cross-L structure can be projected onto orthogonal states along the shared edge. Thus, this structure yields an orthogonal biseparable basis for the three-qubit system. The states in such a basis can be expressed as:
\begin{figure}[t]
    \centering
\includegraphics[scale=.27]{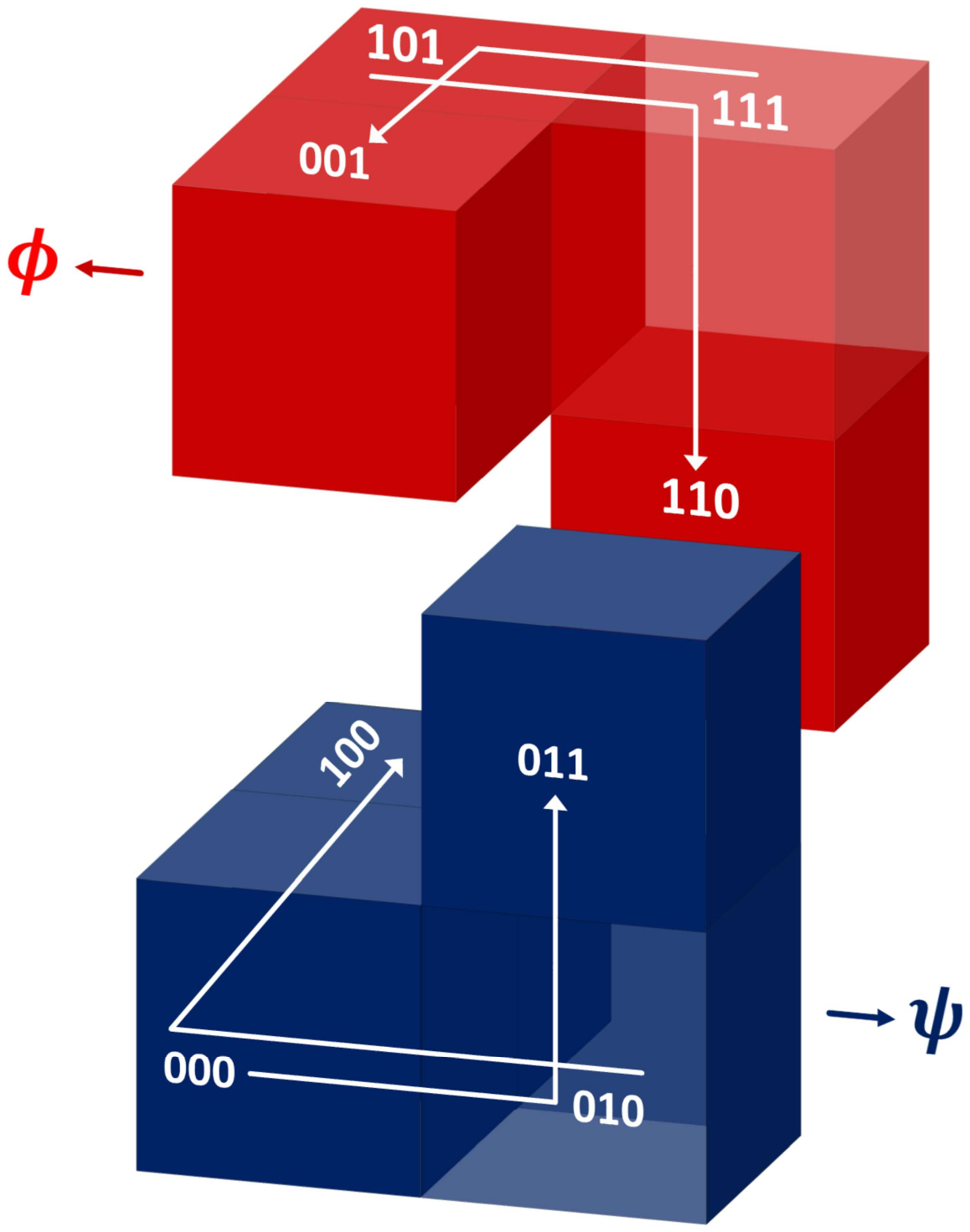}
     \caption{Cross-L structures of biseparable bases $\mathcal{B}$.}
    \label{tile}
\end{figure}
\begin{equation}
\mathcal{B}:\equiv
\begin{array}{l}
 \ket{\phi\down{0}{\ssty 0}^{\ssty-}}\move{2}= \ket{0}(a_{\ssty 0}\ket{00}+a_{\ssty1}\ket{10}-(a_{\ssty 0}+a_{\ssty1})\ket{11}),\\
 \ket{\phi\down{0}{\ssty 0}^{\ssty+}}\move{2}= \ket{0}(a_{\ssty 0}\ket{00}+a_{\ssty1}\ket{10}+\frac{|a_{\ssty 0}|^2+|a_{\ssty1}|^2}{\bar{a_{\ssty 0}}+\bar{a_{\ssty1}}}\ket{11}),\\
 \ket{\phi\down{0}{\ssty 1}^{\ssty-}}\move{2}= (\bar{a_{\ssty 0}}\ket{01}-\bar{a_{\ssty1}}\ket{00}-(\bar{a_{\ssty 0}}-\bar{a_{\ssty1}})\ket{10})\ket{0},\\
 \ket{\phi\down{0}{\ssty 1}^{\ssty+}}\move{2}= (\bar{a_{\ssty 0}}\ket{01}-\bar{a_{\ssty1}}\ket{00}+\frac{|a_{\ssty 0}|^2+|a_{\ssty1}|^2}{a_{\ssty 0}-a_{\ssty1}}\ket{10})\ket{0},\\
 \ket{\psi\down{1}{\ssty 0}^{\ssty-}}= \ket{1}(b_{\ssty 0}\ket{01}+b_{\ssty 1}\ket{11}-(b_{\ssty 0}+b_{\ssty 1})\ket{10}),\\
  \ket{\psi\down{1}{\ssty 0}^{\ssty+}}= \ket{1}(b_{\ssty 0}\ket{01}+b_{\ssty 1}\ket{11}+\frac{|b_{\ssty 0}|^2+|b_{\ssty 1}|^2}{\bar{b_{\ssty 0}}+\bar{b_{\ssty 1}}}\ket{10}),\\
  \ket{\psi\down{1}{\ssty 1}^{\ssty-}}= (\bar{b_{\ssty 0}}\ket{11}-\bar{b_{\ssty 1}}\ket{10}-(\bar{b_{\ssty 0}}-\bar{b_{\ssty 1}})\ket{00})\ket{1},\\
 \ket{\psi\down{1}{\ssty 1}^{\ssty+}}= (\bar{b_{\ssty 0}}\ket{11}-\bar{b_{\ssty 1}}\ket{10}+\frac{|b_{\ssty 0}|^2+|b_{\ssty 1}|^2}{b_{\ssty 0}-b_{\ssty 1}}\ket{00})\ket{1}
 \end{array}
\label{basis_original}
\end{equation}
where, $a_{\ssty 0,1}$, $b_{\ssty 0,1}\in\mathbb{C}$ with $a_{\ssty 0}^2\neq a_{\ssty 1}^2$ and $b_{\ssty 0}^2\neq b_{\ssty 1}^2$. We assume $a_{\ssty 0},b_{\ssty 0}\neq 0$ in order to generate entangled biseperable states via the Cross-L structure. For the shake of bravity, we may set $a_{\ssty 0},b_{\ssty 0}=1$.\hfill$\sty\blacksquare$\par
\end{construction}

In perfect discrimination tasks, the states under consideration must be mutually orthogonal, since even a global measurement cannot perfectly distinguish nonorthogonal states. Consequently, the measurement operators must be chosen such that the post measurement states also remain mutually orthogonal. Any such measurement is called an orthogonality preserving measurement (OPM).  When such measurements act locally, especially under LOCC protocols, they are referred to as orthogonality preserving local measurements (OPLMs).
	
\begin{definition} 
A set of orthogonal quantum states on $\mathcal{H}=\bigotimes_{i=1}^n \mathcal{H}_i$ with $n \geq 2$ and $\operatorname{dim} \mathcal{H}_i \geq 2, i=1, \ldots, n$, is locally irreducible if it is not possible to eliminate one or more states from the set by OPLMs \cite{Saronath2019}.
\end{definition}

Local irreducibility sufficiently ensures local indistinguishability. A sufficient condition for the irreducibility of a set is that there exist no nontrivial measurements as OPLMs. 


\begin{construction}
If we introduce the well known three qubit stopper $\ket{\tau}= \ket{0+1}_A\ket{0+1}_B\ket{0+1}_C$, it omits the $+$ states from the set $\mathcal{B}$. The resulting set is a indistinguishable by the previous argument. The set can be expressed as:
\begin{equation}
\mathcal{S}:\equiv
\begin{array}{l}
 \ket{\phi\down{0}{\ssty 0}^{\ssty-}}\move{2}= \ket{0}(a_{\ssty 0}\ket{00}+a_{\ssty1}\ket{10}-(a_{\ssty 0}+a_{\ssty1})\ket{11}),\\
 \ket{\phi\down{0}{\ssty 1}^{\ssty-}}\move{2}= (\bar{a_{\ssty 0}}\ket{01}-\bar{a_{\ssty1}}\ket{00}-(\bar{a_{\ssty 0}}-\bar{a_{\ssty1}})\ket{10})\ket{0},\\
 \ket{\psi\down{1}{\ssty 0}^{\ssty-}}= \ket{1}(b_{\ssty 0}\ket{01}+b_{\ssty 1}\ket{11}-(b_{\ssty 0}+b_{\ssty 1})\ket{10}),\\
  \ket{\psi\down{1}{\ssty 1}^{\ssty-}}= (\bar{b_{\ssty 0}}\ket{11}-\bar{b_{\ssty 1}}\ket{10}-(\bar{b_{\ssty 0}}-\bar{b_{\ssty 1}})\ket{00})\ket{1},\\
 \ket{\tau\,}\move{8}= \ket{0+1}\ket{0+1}\ket{0+1}
 \end{array}
\label{set_original}
\end{equation}
\end{construction}


\begin{construction}
We consider a simple case to demonstrate the prementioned aspect of imaginarity. Specifically, let $a_{\ssty 1} = -2$ and $b_{\ssty 1} = \mathrm{z}$, with $\mathrm{z} \in \mathbb{C}$. The resulting set can be expressed as:
\begin{equation}
\move{-35}\mathcal{S}_{\mathrm{z}}:\equiv
\begin{array}{l}
 \ket{\phi\down{0}{\ssty 0}^{\ssty-}}\move{2}= \ket{0}(\ket{00}-2\ket{10}+\ket{11}),\\
 \ket{\phi\down{0}{\ssty 1}^{\ssty-}}\move{2}= (\ket{01}+2\ket{00}-3\ket{10})\ket{0},\\
 \ket{\psi\down{1}{\ssty 0}^{\ssty-}}= \ket{1}(\ket{01}+\mathrm{z}\ket{11}-(1+\mathrm{z})\ket{10}),\\
  \ket{\psi\down{1}{\ssty 1}^{\ssty-}}= (\ket{11}-\bar{\mathrm{z}}\ket{10}-(1-\bar{\mathrm{z}})\ket{00})\ket{1},\\
 \ket{\tau\,}\move{8}= \ket{0+1}\ket{0+1}\ket{0+1}
 \end{array}
\label{set_z}
\end{equation}
\end{construction}

As mentioned before, no state of $\mathcal{S}_\mathrm{z}$ can be eliminated under OPLMs. However, there exist bipartite joint measurements that can eliminate some of the states in case $\mathrm{z}$ is real. Surprisingly, if we allow $\mathrm{z}$ to take complex numbers with $\Im(\mathrm{z})\neq0$, the set remains irreducible under any nonglobal measurements. Such a restriction makes the set exhibit the extreme form of nonlocality.

\begin{definition}
Consider a composite quantum system $\mathcal{H}=\bigotimes_{i=1}^n \mathcal{H}_i$ with $n \geq 3$ and $\dim \mathcal{H}_i \geq 3, i=1, \ldots, n$. A set of orthogonal states is strongly nonlocal if it is locally irreducible in every bipartition \cite{Saronath2019}.
\end{definition}

A sufficient condition for strong nonlocality in an $n$-partite system is the nonexistence of any nontrivial joint measurement as OPM by any $n-1$ local observers. In the following theorem, we demonstrate how the mere introduction of the imaginarity is responsible for exhibiting strong nonlocality in the set $\mathcal{S}_\mathrm{z}$.

\begin{theorem}
No bipartite nontrivial OPM exists for the set $\mathcal{S}_\mathrm{z}$  if and only if $\Im(\mathrm{z})\neq0$.
\end{theorem}

The proof of the theorem is included in the appendix \cite{Supplementary}. However, to visualize the role of imaginarity, consider the vectors $v=
\begin{bmatrix}
	1,\mathrm{z} 
\end{bmatrix}^t$ and $\bar{v}=
\begin{bmatrix}
1,\bar{\mathrm{z}} 
\end{bmatrix}^t$. It is evident that the set $\{v,\bar{v}\}$ spans a $1$-dimensional vector space if and only if $\mathrm{z}$ is a real number; otherwise, it spans a two-dimensional vector space. Analogously, when $\mathrm{z}$ is a non-real complex number, the dimension of the subspace spanned by the reduced feature matrices of $\mathcal{S}_\mathrm{z}$ exceeds that of the real case. Consequently, the complementary space contains only scalar multiples of the identity operator. Therefore, no nontrivial measurements can be a part of an OPLM across any bipartitions. As a result, the set $\mathcal{S}_\mathrm{z}$ exhibits strong nonlocality whenever $\mathrm{z}$ has a non-zero imaginary component. 

In contrast, the condition of strong nonlocality may not hold when $\mathrm{z}$ is real. In particular, the set $\mathcal{S}_0$ does not exhibit strong nonlocality. Nevertheless, one may ask whether modifying the set in other ways could restore strong nonlocality. To address this, we replace the so-called product `stopper'  with a biseparable state $\ket{\kappa}= \ket{0+1}_B\ket{00+01+10-11}_{CA}$. This modification converts two of the `minus' states $\ket{\psi\down{0}{\ssty i}^{\ssty-}}$ in $\mathcal{S}_0$ into their `plus' counterparts $\ket{\psi\down{0}{\ssty i}^{\ssty+}}$. Therefore, the states of the original set $\mathcal{S}_0$ and the modified set, denoted by $\mathcal{U}$, can be expressed as follows:
\begin{equation}
\move{-18}\mathcal{S}_0:\equiv
\begin{array}{l}
 \ket{\phi\down{0}{\ssty 0}^{\ssty-}}\move{2}= \ket{0}(\ket{00}-2\ket{10}+\ket{11}),\\
 \ket{\phi\down{0}{\ssty 1}^{\ssty-}}\move{2}= (\ket{01}+2\ket{00}-3\ket{10})\ket{0},\\
  \ket{\psi\down{1}{\ssty 0}^{\ssty-}}= \ket{1}\ket{01-10},\\\ket{\psi\down{1}{\ssty 1}^{\ssty-}}= \ket{00-11}\ket{1},\\
  \ket{\tau\,}\move{8}= \ket{0+1}\ket{0+1}\ket{0+1}
 \end{array}
\label{set_0}
\end{equation}
\begin{equation}
\mathcal{U}:\equiv
\begin{array}{l}
 \ket{\phi\down{0}{\ssty 0}^{\ssty-}}\move{2}= \ket{0}(\ket{00}-2\ket{10}+\ket{11}),\\
 \ket{\phi\down{0}{\ssty 1}^{\ssty-}}\move{2}= (\ket{01}+2\ket{00}-3\ket{10})\ket{0},\\
  \ket{\psi\down{1}{\ssty 0}^{\ssty+}}= \ket{1}\ket{01+10},\\
  \ket{\psi\down{1}{\ssty 1}^{\ssty+}}= \ket{00+11}\ket{1},\\
\ket{\kappa\,}\move{7}= \ket{0+1}_B\ket{00+01+10-11}_{CA}
 \end{array}
\label{UBB}
\end{equation}
i.e., the local correlation of $\ket{\tau}$ between $A$ and $C$ is replaced by an entangled state, which is sufficient for the set $\mathcal{U}$ to exhibit the strongest form of nonlocality.
\begin{theorem}
The set $\mathcal{U}$ is strongly nonlocal, whereas the set $\mathcal{S}_0$ is not \cite{Supplementary}.
\end{theorem}

Beyond cryptographic implications, this result provides a theoretical and operational justification for regarding imaginarity as a fundamental quantum resource, placing it on par with entanglement and coherence in the hierarchy of quantum advantages. In particular, we show how entangling correlations between two distant parties can dilute the effect of imaginarity, and conversely, how imaginarity itself can mimic the role of entanglement. These insights suggest that the presence of complex phases should be understood as a nonclassical resource in its own right, with far-reaching consequences for the study of quantum nonlocality and secure information processing.



The next part of the work mostly includes the application part. We demonstrate how our construction gives rise to several notable advances in quantum information theory. To this end, we begin with a few preliminary definitions that will be required in the subsequent discussion.
\begin{definition}
A non-trivial subspace $V\subset \mathcal{H}= \mathcal{H}_1\bigotimes\cdots\bigotimes \mathcal{H}_k$  is said to be completely entangled subspace, if  $V$ does not contain any nonzero product vector of the form $\ket{u_1}\otimes\cdots\otimes \ket{u_k}$ where $\ket{u_j} \in \mathcal{H}_j, j=1,\ldots,k$. At times, if no confusion can arise, it will be abbreviated as CES.
\end{definition}

\begin{definition}
A non-trivial subspace $V\subset \mathcal{H}$  is said to be genuinely entangled subspace, if  $V$ does not contain any nonzero product vector of the form $\ket{u_1}\otimes\cdots\otimes \ket{u_l}$ where $\ket{u_j} \in \mathcal{K}_j, j=1,\ldots,l$ in any $l$-partition $\mathcal{K}_1\bigotimes\cdots\bigotimes\mathcal{K}_l=\mathcal{H}$. We shall refer to it as GES.
\end{definition}

In the previous paper \cite{Subrata2025}, we developed a sufficient condition for a subspace to be entangled. For that, we go through matrix analysis of product-forming matrices defined for the set of bipartite states (see Eqs. (1,2) in \cite{Subrata2025}).

\begin{lemma}
Let $S=\{\ket{\psi\down{1}{i}}\}\down{1}{\ssty i=0}\up{1}{n-1}$ be a set of $n$ bipartite pure states, with $\{P_m\}$ denoting the collection of all corresponding product forming matrices. The set spans a CES if and only if the system of quadratic equations $X^tP_m X=0$ (for all $m$) admits only the trivial solution $X=0$, where $X\in \mathbb{C}^n$.
\label{product_condition}
\end{lemma}

In this paper, we introduce an algebraic approach to establish a necessary and sufficient condition of CES for improving the visibility of the above lemma. Specifically, we make use of the Gröbner basis formalism, a particular kind of generating set of an ideal in a polynomial ring $K[x_{1},\ldots,x_{n}]$ over a field $K$. A Gröbner basis enables the easy deduction of many important properties of the ideal and the associated algebraic variety, such as the dimension and the number of zeros when it is finite.  Gröbner basis computation is one of the main practical tools for solving systems of polynomial equations and computing the images of algebraic varieties under projections or rational maps. Gröbner basis computation can be seen as a multivariate, non-linear generalization of both Euclid's algorithm for computing polynomial greatest common divisors and Gaussian elimination for linear systems. In the usual case of rational coefficients, this algebraically closed field is chosen as the complex field.

Any set of polynomials may be viewed as a system of polynomial equations by equating the polynomials to zero. The set of solutions of such a system depends only on the generated ideal, and, therefore, does not change when the given generating set is replaced by the Gröbner basis, for any ordering, of the generated ideal. Such a solution, with coordinates in an algebraically closed field containing the coefficients of the polynomials, is called a zero of the ideal. Then, according to Hilbert's Nullstellensatz:

\begin{lemma}
An ideal does not have any zero if and only if its Gröbner basis contains 1.
\label{ideal}
\end{lemma}

Let $S_k$ be the proper subset of $S$ such that $S_k=S\setminus \{\ket{\psi\down{1}{k}}\}$. For a column vector $X\in \mathbb{C}^n$ and $k \in [n]$, define the perturbed vector $X\down{1.5}{k}\in \mathbb{C}^n$ by replacing $k$-th entry of $X$ with $1$.

\begin{theorem}
The set $S$ spans a CES if and only if there exists some $k\in [n]$ such that the subset $S_k$ spans a CES and the Gröbner basis of the ideal generated by the quadratic polynomials $X_k\up{1}{t}P_m X\down{1.5}{k}$ (for all $m$) contains the constant polynomial $1$ \cite{Supplementary}.
\label{Groebner_1}
\end{theorem}

\begin{proposition}
The orthogonal complement of the set $\mathcal{U}$ constitutes a GES in $2\bigotimes2\bigotimes2$ Hilbert space \cite{Supplementary}.
\label{GES}
\end{proposition}

As a direct consequence, there does not exist any pure biseparable state that is orthogonal to the set $\mathcal{U}$. It follows that $\mathcal{U}$ forms a Unextendible Biseparable Basis (UBB) of dimension $5$, thereby providing an explicit construction of a minimal UBB in the three-qubit system. This construction resolves the open problem by confirming that the minimal UBB dimension in $d^{\otimes 3}$ is $d^2 + d - 1$ as proposed in \cite{Agrawal2019}. Moreover, our construction has direct implications for entanglement distillation. In particular, any state supported on the GES admits one-shot distillation.

\begin{proposition}
The orthogonal complement of the set  $\mathcal{U}$ is distillable across every bipartition \cite{Supplementary}.
\end{proposition}

Beyond addressing this foundational question, such constructions of UBB are of independent interest due to their relevance in several recent developments in quantum information science \cite{Saronath2025, Ritabrata2024}. For instance, the authors in \cite{Saronath2025} introduced the concept of splitting a composite Hilbert space into a direct sum of several entangled subspaces.

\begin{definition}
Let $\mathcal{H}$ be a finite-dimensional Hilbert space. A collection of CESs $\{V_1,\ldots, V_m\}$ is called a $m$-CES splitting of $\mathcal{H}$, if $\mathcal{H} = V_1 \oplus\cdots\oplus V_m$. Such a splitting excludes all product states from the Hilbert space.
\end{definition}

\begin{figure}[t]
    \centering
    \includegraphics[scale=1]{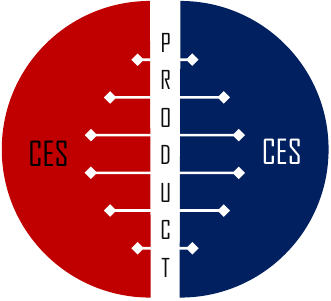}
     \caption{2-CES splitting of Hilbert space.}
    \label{CES}
\end{figure}

In this paper, we mainly investigate $2$-CES splitting due to its significance in the study of mixed state discrimination \cite{Saronath2024, Saronath2025}, entanglement creation from product state\cite{Saronath2025}, and non-entangling perturbations\cite{Ritabrata2024}. It is well known that any two orthogonal pure states are locally distinguishable \cite{Walgate2000}, but this is not necessarily the case when one of the states is mixed \cite{Somshubhro2011, Saronath2024, Saronath2025}. In fact, if we consider an ensemble consisting of a pure state from one half of the 2-CES split and a full-rank mixed state from the other half, the ensemble cannot be perfectly distinguished by LOCC, even in the many-copy scenario. Moreover, two full-rank mixed states supported on each half cannot be unambiguously identified by LOCC with a nonzero probability. On the other hand, projection operators supported on each half of the $2$-CES split can generate entanglement from any product state.

In \cite{Ritabrata2024}, the authors studied the effect of non-entangling perturbations on various subspaces by taking linear spans with specified product vectors. They consider the perturbations on the orthogonal complement of unextendible product bases and also Parthasarathy’s completely entangled spaces, in both bipartite and multipartite scenarios. Certain subspaces preserve the number of product states under such perturbations; these are referred to as stable subspaces.

\begin{definition}
A subspace $V \subset \mathcal{H}$ is said to be \emph{stable} in $\mathcal{H}$ if, for every product state $\ket{\phi} \in V^\perp\subset \mathcal{H}$, the subspace $\ket{\phi} +V$ contains same set of product states as $V$ itself, except for $\ket{\phi}$. If no product state exists in $V^\perp$, then $V$ is called trivially stable.
\end{definition}

Each half of the 2-CES splitting is stable, even when considered within the extended Hilbert space \cite{Extended}, due to the absence of product states from the original Hilbert space. 2-CES splitting does not exist in any $2 \bigotimes n$ system with $n\geqslant 2$. In the bipartite setting, the smallest dimension admitting a $2$-CES splitting is $3 \bigotimes 4$. In the multipartite scenario, however, $2 \bigotimes 2 \bigotimes 2$ system constitutes the smallest Hilbert space that admits $2$-CES splitting. 

\begin{proposition}
Let $W \subset \mathcal{H}^{2\bigotimes2\bigotimes2}$ be the subspace spanned by any four states of the set $\mathcal{U}$. Then the pair $\{W, W^\perp\}$ forms a 2-CES splitting of $\mathcal{H}$.
\label{W_CES}
\end{proposition}

As a consequence, each $W$ and $W^\perp$ is stable under non-entangling perturbation, attaining the maximal possible dimension of CESs in the three-qubit system. However, not every CES exhibits this property. For example, the shift UPB is a well-known case in which the complementary subspace yields a CES that is not stable under non-entangling perturbations \cite{Ritabrata2024}. From this perspective, our constructed UBB $\mathcal{U}$ acquires additional significance by virtue of this stability property. Indeed, the unextendibility of $\mathcal{U}$ guarantees the stability of its span; the key point, however, is to establish the stability of its orthogonal complement. This, in turn, motivates a closer study of the product states contained within the span of $\mathcal{U}$. Although the set $\mathcal{U}$ essentially spans a CES, it contains finitely many pure product states.


\begin{definition}
A non-trivial subspace $V\subset \mathcal{H}= \mathcal{H}_1\bigotimes\cdots\bigotimes \mathcal{H}_k$  is said to be quasi completely entangled subspace, $V$ contain finitely many product vector of the form $\ket{u_1}\otimes\cdots\otimes \ket{u_k}$ where $\ket{u_j} \in \mathcal{H}_j, j=1,\ldots,k$. Hereafter, it will be denoted as QCES. We call the number of product vectors in $V$ as its product index. 
\end{definition}

\begin{lemma}
The set $\mathcal{U}$ spans a QCES with product index $6$.
\label{QCES-6}
\end{lemma}

It is well known that the maximum dimension of a CES in the three-qubit Hilbert space is four \cite{Parthasarathy2004}. Since the set $\mathcal{U}$ spans a five-dimensional subspace, it must necessarily contain some product states. In the appendix, we explicitly identify six product states contained within this subspace \cite{Supplementary}. Notably, however, these six product states prove incapable of perturbing the orthogonal complement of the set $\mathcal{U}$.


%


\begin{proposition}
The orthogonal complement of the subspace spanned by $\mathcal{U}$ is stable even in the extended Hilbert space \cite{Supplementary}.
\end{proposition}

In \cite{Duan2010}, Duan \textit{et al.} proved that the subspace spanned by the shift UPB is locally indistinguishable. Specifically, any orthogonal basis drawn from this subspace cannot be perfectly distinguished by LOCC. We show that our UBB $\mathcal{U}$ similarly spans an indistinguishable subspace.


\begin{proposition}
The subspace spanned by  $\mathcal{U}$ forms an indistinguishable subspace.
\end{proposition}
The QCES spanned by $\mathcal{U}$ contains exactly $6$ distinct product states, none of which are mutually orthogonal \cite{Supplementary}. Consequently, any orthogonal basis $\rho_1,\ldots,\rho_5$ of the QCES can contain at most one rank one state. Therefore, the sum of the orthogonal Schmidt ranks satisfies $\sum_{k=1}^{5}Sch_{\perp}\{\rho_k\}\geqslant4\cdot2+1=9$, which violates the Lemma in \cite{Duan2010}. Hence, the proposition follows.

In conclusion, we have shown that the presence of imaginary components in quantum states provides a distinct operational advantage in multipartite systems. In particular, this advantage manifests as a strengthening of nonlocality within sets of orthogonal pure states from the perspective of state discrimination. We construct a family of orthogonal three-qubit states, which exhibit strong nonlocality if and only if the set involves imaginary components. This establishes imaginarity as a genuine quantum resource with direct cryptographic applications, since information encoded in such states remains secure against a broader class of attacks than previously recognized. 

Our results underscore the indispensable role of complex numbers in quantum theory and open new directions for exploring how imaginarity can be harnessed in quantum communication, nonlocality-based protocols, and resource-theoretic frameworks. However, we construct an unextendible biseparable basis by replacing the product states with a biseparable state whose shared entanglement between two parties nullifies the effect of imaginarity in exhibiting strong nonlocality. This yields the smallest possible set that resolves the open problem regarding the existence of a UBB of cardinality $d^2+d-1$ in $d^{\otimes 3}$ \cite{Agrawal2019}. 

Our construction produces a highly powerful set, rich in resources from multiple perspectives of quantum information theory, and manifesting several fundamental forms of nonlocality. Firstly, any subset of four states splits the Hilbert space into two completely entangled subspaces, each of which remains robust under non-entangling perturbations. Such splits are particularly useful for generating global projective measurements capable of creating entanglement from any product states. Additionally, no ensemble of two full-rank mixed states supported on either half of the split can be unambiguously identified under LOCC. Moreover, both the span of the UBB and its complement remain stable under nontrivial perturbations. Finally, while the UBB spans a locally indistinguishable subspace, its complement, on the other hand, spans a genuinely entangled space which is distilable across every bipartition. This “all-in-one” construction, therefore, offers a versatile and valuable framework with significant potential for applications in quantum information research.

\noindent\textbf{\textit{Acknowledgement}}---The author S. Bera acknowledges the support from CSIR, India. The author I. Biswas acknowledges the support from UGC, India. The authors I. Chattopadhyay and D. Sarkar acknowledge the DST-FIST India.

\bibliography{UBB_5.bib}

\section{Appendix}
\noindent 
The appendix includes explicit calculations and results that support the actual paper to be understood. 

\section{I. P\lowercase{relemenaries}}

Consider a set of $n$ pure states $\{\psi\down{1}{\ssty i}\}\down{1}{\ssty i=0}\up{1}{\ssty n-1}$ on $\mathbb{C}^{d_{\ssty A}}\bigotimes\mathbb{C}^{d_{\ssty B}}$, each of the form 
\begin{equation}
	\ket{\psi\down{1}{i}}_{\move{-1.5}\ssty AB}=\sum_{\ssty k=0}^{d_{\zoom{.5}{A}}\ssty-1}\;\sum_{\ssty l=0}^{d_{\zoom{.5}{B}}\ssty-1} a_{\ssty kl}^{\ssty (i)}\ket{kl}_{\move{-1.5}\ssty AB},\quad a_{\ssty kl}^{\ssty (i)}\in \mathbb{C},
\label{general_state}
\end{equation}
where $a_{\ssty kl}^{\ssty (i)}$ denotes the (k,l)-th $a$-coefficients of the state $\ket{\psi\down{1}{i}}$.

\textbf{\textit{Product Forming Matrix ($n\cross n$):}} 
For $k<p\in [d_{\ssty  A}]$ and $l<r\in [d_{\ssty B}]$, the $(i,j)$-th element of the product forming matrix can be expressed as:
\begin{equation}
\begin{array}{ll}
	(\Lambda_{\ssty kp|lr}^{\ssty    A|B})\down{2}{\ssty i,j} 
	&=\begin{vmatrix}
a_{\ssty kl}^{\ssty(i)} & a_{\ssty kr}^{\ssty(i)} \\[3pt]
a_{\ssty pl}^{\ssty(j)} & a_{\ssty pr}^{\ssty (j)}
\end{vmatrix}\\[14pt]
&
		=a_{\ssty kl}^{\ssty(i)}a_{\ssty pr}^{\ssty (j)}-a_{\ssty kr}^{\ssty(i)}a_{\ssty pl}^{\ssty(j)}
		\end{array}
		\label{product_forming_matrices}
	\end{equation}
		
\textbf{\textit{Reduced Feature Matrix ($d_{\ssty A}\cross d_{\ssty A}$):}} 
For $i\neq j\in[n]$, the reduced feature matrix over $A$ can be defined as:
\begin{equation}
		\Pi_{\ssty ij}^{\ssty    A}=\Tr\down{2}{\ssty \overline{A}}\left(\ketbra{\psi\down{1}{i}}{\psi\down{0}{j}}\right)
\end{equation}

\textbf{\textit{Vectorization:}}
The vectorization of a matrix is a linear transformation that converts the matrix into a vector. Specifically, the vectorization of a $m \times n$ matrix $A$, denoted $\operatorname{vec}(A)$, is the $m n \times 1$ column vector obtained by stacking the columns of the matrix $A$ on top of one another:
$$
\operatorname{vec}(A)=\left[a_{\ssty11}, \ldots, a_{m\ssty1}, a_{\ssty12}, \ldots, a_{m\ssty2}, \ldots, a_{{\ssty1}n}, \ldots, a_{mn}\right]^{\mathrm{T}}
$$ where, $a_{\ssty ij}$ represents the $(i,j)$-th element of $A$, and the superscript ${ }^{\mathrm{T}}$ denotes the transpose. 

\textbf{\textit{Columnwise Vectorization Map:}}
Let $A_{1}, A_2,\ldots,\text{and }A_k$ be $k$ matrices of order $m \times n$ and M be the matrix of order $mn \times k$ whose $i$-th column is the vectorization of $A_i,i=1,2,\ldots,k$. Then, vectorization expresses, through coordinates, the isomorphism $\mathbf{C}^{m \times n}:=\mathbf{C}^m \otimes \mathbf{C}^n \cong \mathbf{C}^{m n}$ between the vector space $V$ spanned by the matrices $\{A_i\}_{i}$ and the column space of $M$. Therefore, the dimension of $V$ is equal to the rank of $M$. The matrix $M$ is defined to be the columnwise vectorization map of the matrices $\{A_i\}_{i}$.


\textbf{\textit{Lemma:}}
In a discrimination task of orthogonal bipartite pure states, a party with local dimension $d$ can go first with a nontrivial OPLM if and only if all the reduced feature matrices over that party span the subspace of dimension $t$, where  $t<d^2-1$.

The contrapositive of this lemma naturally yields a no-go condition within our framework. In particular, the first move is impossible for Alice if and only if the reduced feature matrices over Alice span the entire traceless operator space with dimension $d^2_{\ssty A}-1$. When this no-go condition holds irrespective of the party choices, no party can perform a nontrivial measurement. As a result, no state can be eliminated under OPLM, making the set locally irreducible. Accordingly, a set of orthogonal multipartite states is classified as a strongly nonlocal set if this no-go condition is satisfied across every bipartition. 

We refer to this approach as the reduced feature analysis. There are $n^2-n$ such reduced feature matrices over any subsystem. But it is enough to compute only half of them, since,
\begin{equation}
		\Pi_{\ssty ij}^{\dagger}=\Pi_{\ssty ji},\quad\forall i\neq j\in[n].
\end{equation}
 
 Therfore, for the analysis we compute only $\Pi_{\ssty ij},\forall i<j$, total of $\frac{n^2-n}{2}$ matrices and compute the dimension of the subspace spanned by $\{\Pi_{\ssty ij},\Pi_{\ssty ij}^{\dagger}\}\down{1.5}{\ssty i<j}$.

\textbf{\textit{Entangled Subspace:}} 
A non-trivial subspace $V\subset \mathcal{H}= \mathcal{H}_0\bigotimes\cdots\bigotimes \mathcal{H}_{n-1}$  is called entangled, if it contains no nonzero product vector of the form $\ket{u_{\ssty 0}}\otimes\cdots\otimes \ket{u_{\ssty n-1}}$ where $\ket{u_{\ssty i}} \in \mathcal{H}_i, i\in [n]$.

\textbf{\textit{Biseparable State:}}
An $n$-partite pure state $\ket{\psi}\in \mathcal{H}$  is called biseparable, if  it can be written as a product vector of the form $\ket{u_{\ssty 1}}\otimes\ket{u_{\ssty2}}$ where $\ket{u_{\ssty i}} \in \mathcal{K}_i, j=1,2$ in at least one bipartition $\mathcal{K}_1\bigotimes\mathcal{K}_2=\mathcal{H}$. Otherwise, the state is called genuinely entangled.

\textbf{\textit{Genuinely Entangled Subspace:}}
A non-trivial subspace $V\subset \mathcal{H}$  is called genuinely entangled, if it contains no biseparable state in any bipartition of $\mathcal{H}$. Equivalently, $V$ is genuinely entangled if and only if it is entangled in every bipartition.

\section{II. P\lowercase{roof of} T\lowercase{heorem} 1}

\textbf{\textit{Theorem:}}
No bipartite nontrivial OPM exists for the set $\mathcal{S}_\mathrm{z}$  if and only if $\Im(\mathrm{z})\neq0$.

Case 1:
First, consider the case $\mathrm{z}\in\mathbb{R}$. Suppose,
\begin{equation*}
\zoom{c}{
E=\begin{bmatrix}
			4\mathrm{z}^2 - 4\mathrm{z} & -\mathrm{z}^2 + 2\mathrm{z} + 2 & -\mathrm{z}^3 + 4\mathrm{z}^2 - 2\mathrm{z} + 2 & -\mathrm{z}^3 + 3\mathrm{z}^2 + 3\mathrm{z} - 2 \\
			-\mathrm{z}^2 + 2\mathrm{z} + 2 & -2\mathrm{z}^3 + 4\mathrm{z}^2 + 2\mathrm{z} - 4 & 2\mathrm{z}^2 - \mathrm{z} + 2 & 5\mathrm{z}^2 - 4\mathrm{z} + 2 \\
			-\mathrm{z}^3 + 4\mathrm{z}^2 - 2\mathrm{z} + 2 & 2\mathrm{z}^2 - \mathrm{z} + 2 & 2\mathrm{z}^2 + 2\mathrm{z} - 4 & -\mathrm{z}^3 + 2\mathrm{z}^2 + 2 \\
			-\mathrm{z}^3 + 3\mathrm{z}^2 + 3\mathrm{z} - 2 & 5\mathrm{z}^2 - 4\mathrm{z} + 2 & -\mathrm{z}^3 + 2\mathrm{z}^2 + 2 & 0
		\end{bmatrix}
}
\end{equation*}

It is straightforward to check
\begin{equation}
\bra{\phi}\mathbb{I}\otimes E\ket{\psi}=0,\quad\forall\ket{\phi}\neq\ket{\psi}\in \mathcal{S}_\mathrm{z}
\label{postmeasurement_condition_BC}
\end{equation}

Since the states of $\mathcal{S}_\mathrm{z}$ are mutually orthogonal,
\begin{equation}
\bra{\phi}\ket{\psi}=0,\quad\forall\ket{\phi}\neq\ket{\psi}\in \mathcal{S}_\mathrm{z}
\label{orthogonal_phi_psi}
\end{equation}

The eigenvalues of $E$ are $\lambda_0$,$\lambda_0$,$\lambda_1$, and $\lambda_2$, where
\begin{equation}
	\begin{array}{l}
		\lambda_0 = \mathrm{z}^3 - \mathrm{z}^2 + 3\mathrm{z} - 6, \\
		\lambda_1 = -2\mathrm{z}^3 + 10\mathrm{z}^2 - \mathrm{z} + 2, \\
		\lambda_2 = -2\mathrm{z}^3 + 2\mathrm{z}^2 - 5\mathrm{z} + 2
	\end{array}
\end{equation}

Let us consider $\mu = \max\left(\lambda_0, \lambda_1, \lambda_2\right)$ and $\nu =\max\left( \left| \lambda_0 \right|, \left| \lambda_1 \right|, \left| \lambda_2 \right| \right)$. Define,
\begin{equation}
	M_0 = \dfrac{\mu \mathbb{I}+E}{\mu+\nu},\quad
	M_1 = \dfrac{\nu \mathbb{I}-E}{\mu+\nu}
	\label{genaral_measurement}
\end{equation}

Therefore, for any two arbitrary states $\ket{\phi},\ket{\psi}\in \mathcal{S}_\mathrm{z}$,
$$
\begin{array}{ll}\bra{\phi}\mathbb{I}\otimes M_0\ket{\psi}
&=\dfrac{1}{\mu+\nu}\Bigl(\mu\braket{\phi}{\psi}+\bra{\phi}\mathbb{I}\otimes E\ket{\psi}\Bigl)\\[10pt]
&=0,\quad\text{[by (\ref{orthogonal_phi_psi}) and (\ref{postmeasurement_condition_BC})]}
\end{array}
$$
and,
$$
\begin{array}{ll}
\bra{\phi}\mathbb{I}\otimes M_1\ket{\psi}
&=\dfrac{1}{\mu+\nu}\Bigl(\nu\braket{\phi}{\psi}-\bra{\phi}\mathbb{I}\otimes E\ket{\psi}\Bigl)\\[10pt]
&=0,\quad\text{[by (\ref{orthogonal_phi_psi}) and (\ref{postmeasurement_condition_BC})]}
\end{array}
$$

By definition both $M_0$ and $M_1$ are positive semidefinite operator with the completeness property 
\begin{equation}
M_0+M_1=\mathbb{I}
\end{equation}

Therefore, $\{M_0,M_1\}$ is a valid nontrivial OPM that Bob and Charlie can perform jointly.

Case 2: 
Now we consider $\mathrm{z}$ to be a nonreal complex number. Then, we approach the reduced feature analysis for each of the joint subsystems $AB$, $BC$, and $CA$, each of dimension $4$. 

The reduced feature matrices of $\mathcal{S}_\mathrm{z}$ over the joint subsystem $AB$ are
\begin{equation}
\zoom{c}{
\begin{array}{cc}
			\Pi_{\ssty 01}^{\ssty    AB}=
			\begin{bmatrix}
  \move{8}2 & \move{8}1 & \;-3 & \;0 & \\
  -4 & -2 & \move{8}\;6 & \;0 & \\
  \move{8}0 & \move{8}0 & \move{8}\;0 & \;0 & \\
  \move{8}0 & \move{8}0 & \move{8}\;0 & \;0 & \\
\end{bmatrix},
			&\Pi_{\ssty 02}^{\ssty    AB}=
			\begin{bmatrix}
  \,0 & \;0 & \;0 & \;\;\move{-8}-\mathrm{z} -1 & \\
  \,0 & \;0 & \;1 & \;\;3\mathrm{z} + 2 & \\
  \,0 & \;0 & \;0 & \;\;0 & \\
  \,0 & \;0 & \;0 & \;\;0 & \\
\end{bmatrix},\\[20pt]
			\Pi_{\ssty 03}^{\ssty    AB}=
			\begin{bmatrix}
  0 & 0 & \move{8}\,0 & \;0 & \\
  \mathrm{\bar{z}} - 1 & \,0 & \;-\mathrm{\bar{z}} & 1 & \\
  0 & 0 & \move{8}\,0 & \;0 & \\
  0 & 0 & \move{8}\,0 & \;0 & \\
\end{bmatrix},
			&\Pi_{\ssty 04}^{\ssty    AB}=
			\begin{bmatrix}
  \move{8}1 & \move{8}1 & \move{8}1 & \move{8}\!1 & \\
  -1 & -1 & -1 & \!-1 & \\
  \move{8}0 & \move{8}0 & \move{8}0 & \move{8}\!0 & \\
  \move{8}0 & \move{8}0 & \move{8}0 & \move{8}\!0 & \\
\end{bmatrix},\\[20pt]
			\Pi_{\ssty 12}^{\ssty    AB}=
			\begin{bmatrix}
  0 & 0 & 0 & -2\mathrm{z} - 2 & \\
  0 & 0 & 0 & -\mathrm{z} - 1 & \\
  0 & 0 & 0 & \move{8}3\mathrm{z} + 3 & \\
  0 & 0 & 0 & \move{10}0 & \\
\end{bmatrix},
			&\Pi_{\ssty 13}^{\ssty    AB}=
			\begin{bmatrix}
  &0 & \;\;\;\;0 & \;\;\;\;0 & \;\;\;0 & \\
  &0 & \;\;\;\;0 & \;\;\;\;0 & \;\;\;0 & \\
  &0 & \;\;\;\;0 & \;\;\;\;0 & \;\;\;0 & \\
  &0 & \;\;\;\;0 & \;\;\;\;0 & \;\;\;0 & \\
\end{bmatrix},\\[20pt]
			\Pi_{\ssty 14}^{\ssty    AB}=
			\begin{bmatrix}
  \move{8}2 & \move{8}\!2 & \move{8}\!2 & \move{8}\!2 & \\
  \move{8}1 & \move{8}\!1 & \move{8}\!1 & \move{8}\!1 & \\
  -3 & \!-3 & \!-3 & \!-3 & \\
  \move{8}0 & \move{8}\!0 & \move{8}\!0 & \move{8}\!0 & \\
\end{bmatrix},
			&\Pi_{\ssty 23}^{\ssty    AB}=
			\begin{bmatrix}
  0 & 0 & \move{8}0 & 0 & \\
  0 & 0 & \move{8}0 & 0 & \\
  \mathrm{\bar{z}} - 1 & 0 & -\mathrm{\bar{z}} & 1 & \\
  \mathrm{\bar{z}}^2-\mathrm{\bar{z}} & 0 & -\mathrm{\bar{z}}^2 & \mathrm{\bar{z}} & \\
\end{bmatrix},\\[20pt]
			\Pi_{\ssty 24}^{\ssty    AB}=
			\begin{bmatrix}
  \move{8}0 & \move{8}\!0 & \move{8}\!0 & \move{8}\!0 & \\
  \move{8}0 & \move{8}\!0 & \move{8}\!0 & \move{8}\!0 & \\
  \move{8}1 & \move{8}\!1 & \move{8}\!1 & \move{8}\!1 & \\
  -1 & \!-1 & \!-1 & \!-1 & \\
\end{bmatrix},
			&\Pi_{\ssty 34}^{\ssty    AB}=
			\begin{bmatrix}
  \mathrm{z} - 1 & \move{8}\mathrm{z} - 1 & \move{8}\mathrm{z} - 1 & \move{8}\mathrm{z} - 1  \\
  0 & \move{8}0 & \move{8}0 & \move{8}0  \\
  \move{-8}-\mathrm{z} & -\mathrm{z} & -\mathrm{z} & -\mathrm{z}  \\
  1 & \move{8}1 & \move{8}1 & \move{8}1  \\
\end{bmatrix},\\[20pt]
\end{array}
}
\label{reduced_feature_matrices_S_z_AB}
\end{equation}

The reduced feature matrices of $\mathcal{S}_\mathrm{z}$ over the joint subsystem $BC$ are
\begin{equation}
\zoom{c}{
\begin{array}{cc}
			\Pi_{\ssty 01}^{\ssty    BC}=
			\begin{bmatrix}
  &\move{8}2 & \;\;\;0 & \;\;\;\move{8}1 & \;\;0 & \\
  &\move{8}0 & \;\;\;0 & \;\;\;\move{8}0 & \;\;0 & \\
  &-4 & \;\;\;0 & \;\;\;-2 & \;\;0 & \\
  &\move{8}2 & \;\;\;0 & \;\;\;\move{8}1 & \;\;0 & \\
\end{bmatrix},
			&\Pi_{\ssty 02}^{\ssty    BC}=
			\begin{bmatrix}
  &\;0 & \;\;\;\;0 & \;\;\;\;0 & \;\;\;\;0 & \\
  &\;0 & \;\;\;\;0 & \;\;\;\;0 & \;\;\;\;0 & \\
  &\;0 & \;\;\;\;0 & \;\;\;\;0 & \;\;\;\;0 & \\
  &\;0 & \;\;\;\;0 & \;\;\;\;0 & \;\;\;\;0 & \\
\end{bmatrix},\\[20pt]
			\Pi_{\ssty 03}^{\ssty    BC}=
			\begin{bmatrix}
  &0 & \;\;\bar{\mathrm{z}}- 1 & \;\;0 & \;\;0 & \\
  &0 & \;\;0 & \;\;0 & \;\;0 & \\
  &0 & \;\;2 - 2\bar{\mathrm{z}} & \;\;0 & \;\;0 & \\
  &0 & \;\;\bar{\mathrm{z}} - 1 & \;\;0 & \;\;0 & \\
\end{bmatrix},
			&\Pi_{\ssty 04}^{\ssty    BC}=
			\begin{bmatrix}
  \move{8}1 & \;\move{8}1 & \;\move{8}1 & \move{8}1 & \\
  \move{8}0 & \;\move{8}0 & \;\move{8}0 & \move{8}0 & \\
  -2 & \;-2 & \;-2 & -2 & \\
  \move{8}1 & \;\move{8}1 & \;\move{8}1 & \move{8}1 & \\
\end{bmatrix},\\[20pt]
			\Pi_{\ssty 12}^{\ssty    BC}=
			\begin{bmatrix}
  0 & -3 & 3\mathrm{z} + 3 & -3\mathrm{z} & \\
  0 & \move{8}0 & 0 & \move{8}0 & \\
  0 & \move{8}0 & 0 & \move{8}0 & \\
  0 & \move{8}0 & 0 & \move{8}0 & \\
\end{bmatrix},
			&\Pi_{\ssty 13}^{\ssty    BC}=
			\begin{bmatrix}
  &0 & \;5\bar{\mathrm{z}} - 2 & 0 & -3 & \\
  &0 & \;0 & 0 & \move{8}0 & \\
  &0 & \;\bar{\mathrm{z}} - 1 & 0 & \move{8}0 & \\
  &0 & \;0 & 0 & \move{8}0 & \\
\end{bmatrix},\\[20pt]
			\Pi_{\ssty 14}^{\ssty    BC}=
			\begin{bmatrix}
  &-1 & \;-1 & \,-1 & -1 & \\
  &\move{8}0 & \;\move{8}0 & \,\move{8}0 & \move{8}0 & \\
  &\move{8}1 & \;\move{8}1 & \,\move{8}1 & \move{8}1 & \\
  &\move{8}0 & \;\move{8}0 & \,\move{8}0 & \move{8}0 & \\
\end{bmatrix},
			&\Pi_{\ssty 23}^{\ssty    BC}=
			\begin{bmatrix}
  0 & 0 & 0 & \move{8}0  \\
  0 & \move{-8}-\bar{\mathrm{z}} & 0 & \move{8}1  \\
  0 &\bar{\mathrm{z}}^2 + 1 & 0 & -\bar{\mathrm{z}} - 1  \\
  0 & \move{-4}-\bar{\mathrm{z}}^2 & 0 & \move{8}\bar{\mathrm{z}} \\
\end{bmatrix},\\[20pt]
			\Pi_{\ssty 24}^{\ssty    BC}=
			\begin{bmatrix}
  \move{8}0 & \move{8}0 & \move{8}0 & \move{8}0  \\
  \move{8}1 & \move{8}1 & \move{8}1 & \move{8}1  \\
  -\bar{\mathrm{z}} - 1 & -\bar{\mathrm{z}} - 1 & -\bar{\mathrm{z}} - 1 & -\bar{\mathrm{z}} - 1  \\
  \move{8}\bar{\mathrm{z}} & \move{8}\bar{\mathrm{z}} & \move{8}\bar{\mathrm{z}} & \move{8}\bar{\mathrm{z}}  \\
\end{bmatrix}
,
			&\Pi_{\ssty 34}^{\ssty    BC}=
			\begin{bmatrix}
  \,\move{8}0 & \move{8}0 & \;\move{8}0 & \move{8}0 & \\
  \,-1 & -1 & \;-1 & -1 & \\
  \,\move{8}0 & \move{8}0 & \;\move{8}0 & \move{8}0 & \\
  \,\move{8}1 & \move{8}1 & \;\move{8}1 & \move{8}1 & \\
\end{bmatrix},\\[20pt]
\end{array}
}
\label{reduced_feature_matrices_S_z_BC}
\end{equation}

The reduced feature matrices of $\mathcal{S}_\mathrm{z}$ over the joint subsystem $CA$ are
\begin{equation}
\zoom{c}{
\begin{array}{cc}
			\Pi_{\ssty 01}^{\ssty    CA}=
			\begin{bmatrix}
  &0\; & -3 & \;\;\;\;0 & \;\;\;0 & \\
  &0\; & \move{-2}\;\;\;\;0 & \;\;\;\;0 & \;\;\;0 & \\
  &1\; & \move{-2}\;\;\;\;0 & \;\;\;\;0 & \;\;\;0 & \\
  &0\; & \move{-2}\;\;\;\;0 & \;\;\;\;0 & \;\;\;0 & \\
\end{bmatrix},
			&\Pi_{\ssty 02}^{\ssty    CA}=
			\begin{bmatrix}
  0 & \move{2}2\mathrm{z} + 2 & 0 & 1 - 2\mathrm{z} \\
  0 & \move{2}0 & 0 & 0  \\
  0 & \move{-6}-\mathrm{z} - 1 & 0 & \mathrm{z}  \\
  0 & \move{2}0 & 0 & 0  \\
\end{bmatrix}
,\\[20pt]
			\Pi_{\ssty 03}^{\ssty    CA}=
			\begin{bmatrix}
  0 & 0 & \bar{\mathrm{z}} - 1 & -\bar{\mathrm{z}} - 2 \\
  0 & 0 & 0 & \move{8}0  \\
  0 & 0 & 0 & \move{8}1  \\
  0 & 0 & 0 & \move{8}0  \\
\end{bmatrix},
			&\Pi_{\ssty 04}^{\ssty    CA}=
			\begin{bmatrix}
  -1 & \,-1 & \;-1 & -1 & \\
  \move{8}0 & \,\move{8}0 & \;\move{8}0 & \move{8}0 & \\
  \move{8}1 & \,\move{8}1 & \;\move{8}1 & \move{8}1 & \\
  \move{8}0 & \,\move{8}0 & \;\move{8}0 &\move{8} 0 & \\
\end{bmatrix},\\[20pt]
			\Pi_{\ssty 12}^{\ssty    CA}=
			\begin{bmatrix}
  0 & -\mathrm{z} - 1 & 0 & \mathrm{z} + 2  \\
  0 & 0 & 0 & \move{-4}-3  \\
  0 & 0 & 0 & 0  \\
  0 & 0 & 0 & 0  \\
\end{bmatrix},
			&\Pi_{\ssty 13}^{\ssty    CA}=
			\begin{bmatrix}
  \;0 & 0 & 2\bar{\mathrm{z}} - 2 & 1 - 2\bar{\mathrm{z}}  \\
  \;0 & 0 & 3 - 3\bar{\mathrm{z}} & 3\bar{\mathrm{z}}  \\
  \;0 & 0 & 0 & 0  \\
  \;0 & 0 & 0 & 0  \\
\end{bmatrix}
,\\[20pt]
			\Pi_{\ssty 14}^{\ssty    CA}=
			\begin{bmatrix}
  \move{8}3 & \move{8}3 & \move{8}3 & \move{8}3 & \\
  -3 & -3 & -3 & -3 & \\
  \move{8}0 & \move{8}0 & \move{8}0 & \move{8}0 & \\
  \move{8}0 & \move{8}0 & \move{8}0 & \move{8}0 & \\
\end{bmatrix},
			&\Pi_{\ssty 23}^{\ssty    CA}=
			\begin{bmatrix}
  \;0 & \;0 & 0 & \move{2}0&  \\
  \;0 & \;0 & 0 & \move{-6}-\bar{\mathrm{z}} - 1&  \\
  \;0 & \;0 & 0 & \move{2}0&  \\
  \;0 & \;0 & \bar{\mathrm{z}} - 1 & \move{2}0&  \\
\end{bmatrix},\\[20pt]
			\Pi_{\ssty 24}^{\ssty    CA}=
			\begin{bmatrix}
  \move{8}0 & \move{8}0 & \move{8}0 & \move{8}0  \\
  -\bar{\mathrm{z}} - 1 & -\bar{\mathrm{z}} - 1 & -\bar{\mathrm{z}} - 1 & -\bar{\mathrm{z}} - 1  \\
  \move{8}0 & \move{8}0 & \move{8}0 & \move{8}0  \\
  \move{8}\bar{\mathrm{z}} + 1 & \move{8}\bar{\mathrm{z}} + 1 & \move{8}\bar{\mathrm{z}} + 1 & \move{8}\bar{\mathrm{z}} + 1  \\
\end{bmatrix},
			&\Pi_{\ssty 34}^{\ssty    CA}=
			\begin{bmatrix}
  0 & 0 & 0 & 0  \\
  0 & 0 & 0 & 0  \\
  \mathrm{z} - 1 & \mathrm{z} - 1 & \mathrm{z} - 1 & \mathrm{z} - 1  \\
  1 - \mathrm{z} & 1 - \mathrm{z} & 1 - \mathrm{z} & 1 - \mathrm{z}  \\
\end{bmatrix},\\[20pt]
\end{array}
}
\label{reduced_feature_matrices_S_z_CA}
\end{equation}

In each case, we then evaluate the corresponding vectorization map of $\{\Pi_{\ssty ij},\Pi_{\ssty ij}^{\dagger}\}\down{1.5}{\ssty i<j}$, which produces an $16\cross20$ matrix (computed using our algorithm), and find its rank to be $15$ for each $\Im(\mathrm{z})\neq0$, which is exactly one less than the square of the dimension of the corresponding joint subsystem.

\section{III. P\lowercase{roof of} T\lowercase{heorem} 2}

\textbf{\textit{Theorem:}}
The set $\mathcal{U}$ is strongly nonlocal.

Proof:
Let the states of the set $\mathcal{S}_0$ can be reexpressed as
\begin{equation}
\move{-18}\mathcal{S}_0:\equiv
\begin{array}{l}
\ket{\phi\down{0}{\ssty 0}^{\ssty-}}\move{2}= \ket{0}(\ket{00}-2\ket{10}+\ket{11}),\\
\ket{\phi\down{0}{\ssty 1}^{\ssty-}}\move{2}= (\ket{01}+2\ket{00}-3\ket{10})\ket{0},\\
\ket{\psi\down{1}{\ssty 0}^{\ssty-}}= \ket{1}\ket{01-10},\\
\ket{\psi\down{1}{\ssty 1}^{\ssty-}}= \ket{00-11}\ket{1},\\
\ket{\tau\,}\move{8}= \ket{0+1}\ket{0+1}\ket{0+1}
\end{array}
\end{equation}

For $\mathrm{z}=0$, the measurements (\ref{genaral_measurement}) produces projection operators:
\begin{equation}
	M_0 =\frac{1}{4}
	\begin{bmatrix}
		\move{8}3 & 1 & 1 & \move{-4}-1 \\
		\move{8}1 & 1 & 1 & \move{4}1 \\
		\move{8}1 & 1 & 1 & \move{4}1 \\
		-1 & 1 & 1 & \move{4}3
	\end{bmatrix}\move{-2},\quad
	M_1 =\frac{1}{4}
	\begin{bmatrix}
		\move{8}1 & -1 & -1 & \move{8}1 \\
		-1 & \move{8}3 & -1 & -1 \\
		-1 & -1 & \move{8}3 & -1 \\
		\move{8}1 & -1 & -1 & \move{8}1
	\end{bmatrix}
\end{equation}

The following relations hold: 
\begin{equation}
\bra{\phi\down{0}{\ssty 0}^{\ssty-}}\mathbb{I}\otimes M_0\ket{\phi\down{0}{\ssty 0}^{\ssty-}}=0,\quad \bra{\psi\down{1}{\ssty 0}^{\ssty-}}\mathbb{I}\otimes M_0\ket{\psi\down{1}{\ssty 0}^{\ssty-}}=0,
\end{equation}
\begin{equation}
\bra{\tau\,}\mathbb{I}\otimes M_1\ket{\tau\,}=0
\end{equation}

Therefore, when Bob and Charlie perform the joint measurement $\{M_0,M_1\}$, the outcome $0$ eliminate the states $\ket{\phi\down{0}{\ssty 0}^{\ssty-}}$ and $\ket{\psi\down{1}{\ssty 0}^{\ssty-}}$, while the outcome $1$ eliminate the state $\ket{\tau\,}$. Therefore, the set $\mathcal{S}_0$ does not exhibit strong nonlocality.

\textbf{\textit{Theorem:}}
The set $\mathcal{S}_0$ is not strongly nonlocal.

Proof:
The states of the set $\mathcal{U}$ can be expressed as
\begin{equation}
\mathcal{U}:\equiv
\begin{array}{l}
 \ket{\phi\down{0}{\ssty 0}^{\ssty-}}\move{2}= \ket{0}(\ket{00}-2\ket{10}+\ket{11}),\\
 \ket{\phi\down{0}{\ssty 1}^{\ssty-}}\move{2}= (\ket{01}+2\ket{00}-3\ket{10})\ket{0},\\
  \ket{\psi\down{1}{\ssty 0}^{\ssty+}}= \ket{1}\ket{01+10},\\
  \ket{\psi\down{1}{\ssty 1}^{\ssty+}}= \ket{00+11}\ket{1},\\
\ket{\kappa\,}\move{7}= \ket{0+1}\ket{00+10}+\ket{0-1}\ket{01+11}
 \end{array}
\label{UBB}
\end{equation}

 The reduced feature matrices of $\mathcal{U}$ over the joint subsystem $AB$ are
\begin{equation}
\zoom{c}{
\begin{array}{cc}
			\Pi_{\ssty 01}^{\ssty    AB}=
			\begin{bmatrix}
  \move{8}2 & \move{8}1 & -3 & \move{8}0 & \\
  -4 & -2 & \move{8}6 & \move{8}0 & \\
  \move{8}0 & \move{8}0 & \move{8}0 & \move{8}0 & \\
  \move{8}0 & \move{8}0 & \move{8}0 & \move{8}0 & \\
\end{bmatrix},
			&\Pi_{\ssty 02}^{\ssty    AB}=
			\begin{bmatrix}
  \move{8}0 & \move{8}0 & \move{8}0 & \move{8}1 & \\
  \move{8}0 & \move{8}0 & \move{8}1 & -2 & \\
  \move{8}0 & \move{8}0 & \move{8}0 & \move{8}0 & \\
  \move{8}0 & \move{8}0 & \move{8}0 & \move{8}0 & \\
\end{bmatrix},\\[20pt]
			\Pi_{\ssty 03}^{\ssty    AB}=
			\begin{bmatrix}
  \move{8}0 & \move{8}0 & \move{8}0 & \move{8}0 & \\
 \move{8}1 & \move{8}0 & \move{8}0 & \move{8}1 & \\
  \move{8}0 & \move{8}0 & \move{8}0 & \move{8}0 & \\
  \move{8}0 & \move{8}0 & \move{8}0 & \move{8}0 & \\
\end{bmatrix},
			&\Pi_{\ssty 04}^{\ssty    AB}=
			\begin{bmatrix}
  \move{8}1 & \move{8}1 & \move{8}1 & \move{8}1 & \\
  -1 & -1 & -3 & -3 & \\
  \move{8}0 & \move{8}0 & \move{8}0 & \move{8}0 & \\
  \move{8}0 & \move{8}0 & \move{8}0 & \move{8}0 & \\
\end{bmatrix}
,\\[20pt]
			\Pi_{\ssty 12}^{\ssty    AB}=
			\begin{bmatrix}
  \move{8}0 & \move{8}0 & \move{8}0 & \move{8}2 & \\
  \move{8}0 & \move{8}0 & \move{8}0 & \move{8}1 & \\
  \move{8}0 & \move{8}0 & \move{8}0 & -3 & \\
  \move{8}0 & \move{8}0 & \move{8}0 & \move{8}0 & \\
\end{bmatrix},
			&\Pi_{\ssty 13}^{\ssty    AB}=
			\begin{bmatrix}
  \move{8}0 & \move{8}0 & \move{8}0 & \move{8}0 & \\
  \move{8}0 & \move{8}0 & \move{8}0 & \move{8}0 & \\
  \move{8}0 & \move{8}0 & \move{8}0 & \move{8}0 & \\
  \move{8}0 & \move{8}0 & \move{8}0 & \move{8}0 & \\
\end{bmatrix},\\[20pt]
			\Pi_{\ssty 14}^{\ssty    AB}=
			\begin{bmatrix}
  \move{8}2 & \move{8}2 & \move{8}2 & \move{8}2 & \\
  \move{8}1 & \move{8}1 & \move{8}1 & \move{8}1 & \\
  -3 & -3 & -3 & -3 & \\
  \move{8}0 & \move{8}0 & \move{8}0 & \move{8}0 & \\
\end{bmatrix}
,
			&\Pi_{\ssty 23}^{\ssty    AB}=
			\begin{bmatrix}
  \move{8}0 & \move{8}0 & \move{8}0 & \move{8}0 & \\
  \move{8}0 & \move{8}0 & \move{8}0 & \move{8}0 & \\
  \move{8}1 & \move{8}0 & \move{8}0 & \move{8}1 & \\
  \move{8}0 & \move{8}0 & \move{8}0 & \move{8}0 & \\
\end{bmatrix},\\[20pt]
			\Pi_{\ssty 24}^{\ssty    AB}=
			\begin{bmatrix}
  \move{8}0 & \move{8}0 & \move{8}0 & \move{8}0 & \\
  \move{8}0 & \move{8}0 & \move{8}0 & \move{8}0 & \\
  \move{8}1 & \move{8}1 & -1 & -1 & \\
  \move{8}1 & \move{8}1 & \move{8}1 & \move{8}1 & \\
\end{bmatrix},
			&\Pi_{\ssty 34}^{\ssty    AB}=
			\begin{bmatrix}
  \move{8}1 & \move{8}1 & -1 & -1 & \\
  \move{8}0 & \move{8}0 & \move{8}0 & \move{8}0 & \\
  \move{8}0 & \move{8}0 & \move{8}0 & \move{8}0 & \\
  \move{8}1 & \move{8}1 & -1 & -1 & \\
\end{bmatrix},\\[20pt]
\end{array}
}
\label{reduced_feature_matrices_U_AB}
\end{equation}

The reduced feature matrices of $\mathcal{U}$ over the joint subsystem $BC$ are
\begin{equation}
\zoom{c}{
\begin{array}{cc}
			\Pi_{\ssty 01}^{\ssty    BC}=
			\begin{bmatrix}
  \move{8}2 & \move{8}0 & \move{8}1 & \move{8}0 & \\
  \move{8}0 & \move{8}0 & \move{8}0 & \move{8}0 & \\
  -4 & \move{8}0 & -2 &\move{8} 0 & \\
  \move{8}2 & \move{8}0 & \move{8}1 & \move{8}0 & \\
\end{bmatrix},
			&\Pi_{\ssty 02}^{\ssty    BC}=
			\begin{bmatrix}
  \move{8}0 & \move{8}0 & \move{8}0 & \move{8}0 & \\
  \move{8}0 & \move{8}0 & \move{8}0 & \move{8}0 & \\
  \move{8}0 & \move{8}0 & \move{8}0 & \move{8}0 & \\
  \move{8}0 & \move{8}0 & \move{8}0 & \move{8}0 & \\
\end{bmatrix}
,\\[20pt]
			\Pi_{\ssty 03}^{\ssty    BC}=
			\begin{bmatrix}
  \move{8}0 & \move{8}1 & \move{8}0 & \move{8}0 & \\
  \move{8}0 & \move{8}0 & \move{8}0 & \move{8}0 & \\
  \move{8}0 & -2 & \move{8}0 & \move{8}0 & \\
  \move{8}0 & \move{8}1 & \move{8}0 & \move{8}0 & \\
\end{bmatrix},
			&\Pi_{\ssty 04}^{\ssty    BC}=
			\begin{bmatrix}
  \move{8}1 & \move{8}1 & \move{8}1 & \move{8}1 & \\
  \move{8}0 & \move{8}0 & \move{8}0 & \move{8}0 & \\
  -2 & -2 & -2 & -2 & \\
  \move{8}1 & \move{8}1 & \move{8}1 & \move{8}1 & \\
\end{bmatrix}
,\\[20pt]
			\Pi_{\ssty 12}^{\ssty    BC}=
			\begin{bmatrix}
  \move{8}0 & -3 & -3 & \move{8}0 & \\
  \move{8}0 & \move{8}0 & \move{8}0 & \move{8}0 & \\
  \move{8}0 & \move{8}0 & \move{8}0 & \move{8}0 & \\
  \move{8}0 & \move{8}0 & \move{8}0 & \move{8}0 & \\
\end{bmatrix},
			&\Pi_{\ssty 13}^{\ssty    BC}=
			\begin{bmatrix}
  \move{8}0 & \move{8}2 & \move{8}0 & -3 & \\
  \move{8}0 & \move{8}0 & \move{8}0 & \move{8}0 & \\
  \move{8}0 & \move{8}1 & \move{8}0 & \move{8}0 & \\
  \move{8}0 & \move{8}0 & \move{8}0 & \move{8}0 & \\
\end{bmatrix},\\[20pt]
			\Pi_{\ssty 14}^{\ssty    BC}=
			\begin{bmatrix}
  -1 & \move{8}5 & -1 & \move{8}5 & \\
  \move{8}0 & \move{8}0 & \move{8}0 & \move{8}0 & \\
  \move{8}1 & \move{8}1 & \move{8}1 & \move{8}1 & \\
  \move{8}0 & \move{8}0 & \move{8}0 & \move{8}0 & \\
\end{bmatrix},
			&\Pi_{\ssty 23}^{\ssty    BC}=
			\begin{bmatrix}
  \move{8}0 & \move{8}0 & \move{8}0 & \move{8}0 & \\
  \move{8}0 & \move{8}0 & \move{8}0 & \move{8}1 & \\
  \move{8}0 & \move{8}0 & \move{8}0 & \move{8}1 & \\
  \move{8}0 & \move{8}0 & \move{8}0 & \move{8}0 & \\
\end{bmatrix}
,\\[20pt]
			\Pi_{\ssty 24}^{\ssty    BC}=
			\begin{bmatrix}
  \move{8}0 & \move{8}0 & \move{8}0 & \move{8}0 & \\
  \move{8}1 & -1 & \move{8}1 & -1 & \\
  \move{8}1 & -1 & \move{8}1 & -1 & \\
  \move{8}0 & \move{8}0 & \move{8}0 & \move{8}0 & \\
\end{bmatrix}
,
			&\Pi_{\ssty 34}^{\ssty    BC}=
			\begin{bmatrix}
  \move{8}0 & \move{8}0 & \move{8}0 & \move{8}0 & \\
  \move{8}1 & \move{8}1 & \move{8}1 & \move{8}1 & \\
  \move{8}0 & \move{8}0 & \move{8}0 & \move{8}0 & \\
  \move{8}1 & -1 & \move{8}1 & -1 & \\
\end{bmatrix}
,\\[20pt]
\end{array}
}
\label{reduced_feature_matrices_U_BC}
\end{equation}

The reduced feature matrices of $\mathcal{U}$ over the joint subsystem $CA$ are
\begin{equation}
\zoom{c}{
\begin{array}{cc}
			\Pi_{\ssty 01}^{\ssty    CA}=
			\begin{bmatrix}
  \move{8}0 & -3 & \move{8}0 & \move{8}0 & \\
  \move{8}0 & \move{8}0 & \move{8}0 & \move{8}0 & \\
  \move{8}1 & \move{8}0 & \move{8}0 & \move{8}0 & \\
  \move{8}0 & \move{8}0 & \move{8}0 & \move{8}0 & \\
\end{bmatrix},
			&\Pi_{\ssty 02}^{\ssty    CA}=
			\begin{bmatrix}
  \move{8}0 & -2 & \move{8}0 & \move{8}1 & \\
  \move{8}0 & \move{8}0 & \move{8}0 & \move{8}0 & \\
  \move{8}0 & \move{8}1 & \move{8}0 & \move{8}0 & \\
  \move{8}0 & \move{8}0 & \move{8}0 & \move{8}0 & \\
\end{bmatrix}
,\\[20pt]
			\Pi_{\ssty 03}^{\ssty    CA}=
			\begin{bmatrix}
  \move{8}0 & \move{8}0 & \move{8}1 & -2 & \\
  \move{8}0 & \move{8}0 & \move{8}0 & \move{8}0 & \\
  \move{8}0 & \move{8}0 & \move{8}0 & \move{8}1 & \\
  \move{8}0 & \move{8}0 & \move{8}0 & \move{8}0 & \\
\end{bmatrix},
			&\Pi_{\ssty 04}^{\ssty    CA}=
			\begin{bmatrix}
  -1 & -1 & -1 & \move{8}1 & \\
  \move{8}0 & \move{8}0 & \move{8}0 & \move{8}0 & \\
  \move{8}1 & \move{8}1 & \move{8}1 & -1 & \\
  \move{8}0 & \move{8}0 & \move{8}0 & \move{8}0 & \\
\end{bmatrix},\\[20pt]
			\Pi_{\ssty 12}^{\ssty    CA}=
			\begin{bmatrix}
  \move{8}0 & \move{8}1 & \move{8}0 & \move{8}2 & \\
  \move{8}0 & \move{8}0 & \move{8}0 & -3 & \\
  \move{8}0 & \move{8}0 & \move{8}0 & \move{8}0 & \\
  \move{8}0 & \move{8}0 & \move{8}0 & \move{8}0 & \\
\end{bmatrix},
			&\Pi_{\ssty 13}^{\ssty    CA}=
			\begin{bmatrix}
  \move{8}0 & \move{8}0 & \move{8}2 & \move{8}1 & \\
  \move{8}0 & \move{8}0 & -3 & \move{8}0 & \\
  \move{8}0 & \move{8}0 & \move{8}0 & \move{8}0 & \\
  \move{8}0 & \move{8}0 & \move{8}0 & \move{8}0 & \\
\end{bmatrix}
,\\[20pt]
			\Pi_{\ssty 14}^{\ssty    CA}=
			\begin{bmatrix}
  \move{8}3 & \move{8}3 & \move{8}3 & -3 & \\
  -3 & -3 & -3 & \move{8}3 & \\
  \move{8}0 & \move{8}0 & \move{8}0 & \move{8}0 & \\
  \move{8}0 & \move{8}0 & \move{8}0 & \move{8}0 & \\
\end{bmatrix},
			&\Pi_{\ssty 23}^{\ssty    CA}=
			\begin{bmatrix}
  \move{8}0 & \move{8}0 & \move{8}0 & \move{8}0 & \\
  \move{8}0 & \move{8}0 & \move{8}0 & \move{8}1 & \\
  \move{8}0 & \move{8}0 & \move{8}0 & \move{8}0 & \\
  \move{8}0 & \move{8}0 & \move{8}1 & \move{8}0 & \\
\end{bmatrix},\\[20pt]
			\Pi_{\ssty 24}^{\ssty    CA}=
			\begin{bmatrix}
  \move{8}0 & \move{8}0 & \move{8}0 & \move{8}0 & \\
  \move{8}1 & \move{8}1 & \move{8}1 & -1 & \\
  \move{8}0 & \move{8}0 & \move{8}0 & \move{8}0 & \\
  \move{8}1 & \move{8}1 & \move{8}1 & -1 & \\
\end{bmatrix},
			&\Pi_{\ssty 34}^{\ssty    CA}=
			\begin{bmatrix}
  \move{8}0 & \move{8}0 & \move{8}0 & \move{8}0 & \\
  \move{8}0 & \move{8}0 & \move{8}0 & \move{8}0 & \\
  \move{8}1 & \move{8}1 & \move{8}1 & -1 & \\
  \move{8}1 & \move{8}1 & \move{8}1 & -1 & \\
\end{bmatrix},\\[20pt]
\end{array}
}
\label{reduced_feature_matrices_U_CA}
\end{equation}

In each case, we then evaluate the corresponding vectorization map of $\{\Pi_{\ssty ij},\Pi_{\ssty ij}^{\dagger}\}\down{1.5}{\ssty i<j}$, which produces an $16\cross20$ matrix (computed using our algorithm), and find its rank to be $15$, which is exactly one less than the square of the dimension of the corresponding joint subsystem.

\section{IV. P\lowercase{roof of} L\lowercase{emma} 1}

\textbf{\textit{Lemma:}}
Let $S=\{\ket{\psi\down{1}{i}}\}\down{1}{\ssty i=0}\up{1}{n-1}$ be a set of $n$ bipartite pure states, with $\{P_m\}$ denoting the collection of all corresponding product forming matrices. The set spans a CES if and only if the system of quadratic equations $X^tP_m X=0$ (for all $m$) admits only the trivial solution $X=0$, where $X\in \mathbb{C}^n$.

Proof:
An arbitrary superposition of $n$ pure states can be expressed as:
\begin{equation*}
	\begin{array}{ll}
	\ket{\chi}_{\move{-1.5}\ssty AB}
	&=\sum\limits_{ i=0}^{ n-1} x_{\ssty i}\ket{\psi\down{1}{i}}_{\move{-1.5}\ssty AB},\quad x_{\ssty i}\in \mathbb{C}\\
	&=\sum\limits_{ i=0}^{ n-1}  x_{\ssty i}(\sum\limits_{\ssty k}^{d_{\ssty  A}{\ssty-1}}\;\sum\limits_{\ssty l=0}^{d_{\ssty  B}{\ssty-1}} a_{\ssty kl}^{\ssty (i)}\ket{kl}_{\move{-1.5}\ssty AB})\quad\text{[by (\ref{general_state})]}\\
	&=\sum\limits_{\ssty k=0}^{d_{\ssty  A}{\ssty-1}}\ket{k}_{\move{-1.5}\ssty A}\otimes(\sum\limits_{ i=0}^{ n-1}\;\sum\limits_{\ssty l=0}^{d_{\ssty  B}{\ssty-1}} x_{\ssty i}a_{\ssty l}^{\ssty (i)}\ket{kl}_{\move{-1.5}\ssty B})\\
	&=\sum\limits_{k=0}^{\ssty d_{\ssty A}-1}\ket{k}_{\move{-1.5}\ssty A}\ket{\phi\down{1}{k}}_{\move{-1.5}\ssty B}
	\label{arbitrary_state}
	\end{array}
\end{equation*}
where $\ket{\phi\down{1}{k}}=\sum_{\ssty il}x_{\ssty i} a_{\ssty kl}^{\ssty (i)}\ket{l}\in\mathcal{H}^{\ssty B}$ for $k\in [d_{\ssty A}]$.

The state $\ket{\chi}$ is a product state if and only if any pair of the collection of vectors $\{\ket{\phi\down{1}{k}}\}_{k=0}^{\ssty d_{\ssty A}-1}$ are linearly dependent, i.e., $\ket{\phi\down{1}{k}}$ and $\ket{\phi\down{1}{p}}$ are linearly dependent for each $k,p(>k)\in[d_{\ssty  A}]$. This is equivalent to collinearity between two vectors $(\sum_{\ssty i}x_{\ssty i}a_{\ssty kl}^{\ssty (i)})\ket{l}+(\sum_{\ssty i}x_{\ssty i}a_{\ssty kr}^{\ssty (i)})\ket{ r}$ and $(\sum_{\ssty i}x_{\ssty i}a_{\ssty pl}^{\ssty (i)})\ket{ l}+(\sum_{\ssty i}x_{\ssty i}a_{\ssty pr}^{\ssty (i)})\ket {r}$ for each $l,r(> l)\in[d_{\ssty  B}]$. Therefore, 
\begin{equation}
\begin{array}{ll}
	&\begin{vmatrix}
		\sum\limits_{\ssty i=0}^{\ssty n-1}\,x_{\ssty i} a_{\ssty kl}^{\ssty (i)} & \sum\limits_{\ssty i=0}^{\ssty n-1}\,x_{\ssty i} a_{\ssty kr}^{\ssty (i)} \\[10pt]
		\sum\limits_{\ssty i=0}^{\ssty n-1}\,x_{\ssty i} a_{\ssty pl}^{\ssty (i)} & \sum\limits_{\ssty i=0}^{\ssty n-1}\,x_{\ssty i} a_{\ssty pr}^{\ssty (i)}
	\end{vmatrix}=0 \\
	\Rightarrow &
	(\sum\limits_{\ssty i=0}^{\ssty n-1}\,x_{\ssty i} a_{\ssty kl}^{\ssty (i)})(\sum\limits_{\ssty i=0}^{\ssty n-1}\,x_{\ssty i} a_{\ssty pr}^{\ssty (i)})-(\sum\limits_{\ssty i=0}^{\ssty n-1}\,x_{\ssty i} a_{\ssty kr}^{\ssty (i)})(\sum\limits_{\ssty i=0}^{\ssty n-1}\,x_{\ssty i} a_{\ssty pl}^{\ssty (i)})=0\\
	\Rightarrow &
	 (\sum\limits_{\ssty i=0}^{\ssty n-1}\,x_{\ssty i} a_{\ssty kl}^{\ssty (i)})(\sum\limits_{\ssty j=0}^{\ssty n-1}\,x_{\ssty j} a_{\ssty pr}^{\ssty (j)})-(\sum\limits_{\ssty i=0}^{\ssty n-1}\,x_{\ssty i} a_{\ssty kr}^{\ssty (i)})(\sum\limits_{\ssty j=0}^{\ssty n-1}\,x_{\ssty j} a_{\ssty pl}^{\ssty (j)})=0\\
	\Rightarrow &
	 \sum\limits_{\ssty i=0}^{\ssty n-1}\,\sum\limits_{\ssty j=0}^{\ssty n-1}x_{\ssty i} x_{\ssty j} a_{\ssty kl}^{\ssty (i)}a_{\ssty pr}^{\ssty (j)}-\sum\limits_{\ssty i=0}^{\ssty n-1}\,\sum\limits_{\ssty j=0}^{\ssty n-1}x_{\ssty i} x_{\ssty j} a_{\ssty kr}^{\ssty (i)}a_{\ssty pl}^{\ssty (j)}=0\\
	 \Rightarrow &
	 \sum\limits_{\ssty i=0}^{\ssty n-1}\,\sum\limits_{\ssty j=0}^{\ssty n-1}x_{\ssty i} x_{\ssty j} (a_{\ssty kl}^{\ssty (i)}a_{\ssty pr}^{\ssty (j)}- a_{\ssty kr}^{\ssty (i)}a_{\ssty pl}^{\ssty (j)})=0\\
	 \Rightarrow &
	 \sum\limits_{\ssty i=0}^{\ssty n-1}\,\sum\limits_{\ssty j=0}^{\ssty n-1}x_{\ssty i} x_{\ssty j} (\Lambda_{\ssty kp|lr}^{\ssty   A|B})\down{2}{\ssty i,j}=0,\quad\text{[by (\ref{product_forming_matrices})]}\\
	 \Rightarrow &
	 X^t\Lambda_{\ssty kp|lr}^{\ssty   A|B}\,X=0\\
	 \Rightarrow &
	 X^tP_m X=0
	\end{array}
	\label{2x2_block}
\end{equation}
where $X=(x_0,x_1,\ldots,x_{n-1})^{t}$ is the column vector with $x_i$ denoting its $i$-th entry, and $\Lambda_{\ssty kp|lr}^{\ssty AB}=P_m$ with $m$ running over all choices of the indices $k,p,l,r$.

Every nonzero solution of eqs. (\ref{2x2_block}) corresponds to a product state within the span of $S$. In other words, the system of homogeneous equations $X^tP_m X=0$ (for all $m$) admits no nonzero solution if and only if the span of $S$ consists entirely of entangled states.

\section{V. P\lowercase{roof of} T\lowercase{heorem} 3}

Let $S_k$ denote the set of $n-1$ pure states obtained by removing the $k$-th state from $S=\{\ket{\psi\down{1}{i}}\}\down{1}{\ssty i=0}\up{1}{n-1}$. For the column vector $X\in \mathbb{C}^n$ and $k \in [n]$, define the perturbed vector $X\down{1}{k}\in \mathbb{C}^n$ by replacing $k$-th entry of $X$ with $1$.

\textbf{\textit{Theorem:}}
The set $S$ spans a CES if and only if there exists some $k\in [n]$ such that the subset $S_k$ spans a CES and the Gröbner basis of the ideal generated by the quadratic polynomials $X_k\up{1}{t}P_m X\down{1.5}{k}$ (for all $m$) contains the constant polynomial $1$.

At first, assume that $S$ spans a CES. Therefore, every subset of it must also span a CES, in particular, each $S_k$ for all $k\in [n]$. Since $S$ contains no product state, Lemma \ref{product_condition} implies that the system of quadratic equations (\ref{2x2_block}) does not admit any nonzero solution. Consequently, no vector $X_k$ satisfies Eqs. (\ref{2x2_block}), because $X_k\neq0$ for all $k\in [n]$. Accordingly, the system of quadratic polynomials $\{X_k^tP_m X\down{1.5}{k}\}\down{1.5}{m}$ admits no solution for each $k\in[n]$. By Lemma \ref{ideal}, the associated Gröbner basis necessarily contains the constant polynomial $1$.

Conversely, assume the stated condition holds. Then there exist $k \in [n]$, such the subset $S_k$ spans a CES and the Gröbner basis of $\{X_k^tP_m X\down{1.5}{k}\}\down{1.5}{m}$ contains $1$, which means the system admits no solution. Suppose, for contradiction, that $S$ does not span a CES. Then, for some $a_{\ssty i}\in\mathbb{C}$, there exists a product state of the form 
\begin{equation}
\sum_{ i=0}^{ n-1} a_{\ssty i}\ket{\psi\down{1}{i}}
\end{equation} 

Since $S_k$ spans a CES, the state must satisfy $y_k\neq0$. Consequently, the state 
\begin{equation}
\sum_{ i=0}^{ n-1}\; \frac{a_{\ssty i}}{a\down{1}{\ssty k}}\;\ket{\psi\down{1}{i}}
\end{equation} 
is also a product state. Hence, the vector 
\begin{equation}
\left(\frac{a_{\ssty 0}}{a\down{1}{\ssty k}},\frac{a_{\ssty 1}}{a\down{1}{\ssty k}},\ldots,\frac{a_{\ssty k-1}}{a\down{1}{\ssty k}},1,\frac{a_{\ssty k+1}}{a\down{1}{\ssty k}},\ldots,\frac{a_{\ssty n-1}}{a\down{1}{\ssty k}}\right)^{t}
\end{equation} 
provides a solution of the system $\{X_k^tP_m X\down{1.5}{k}=0\}\down{1.5}{m}$. This contradicts the assumption that $S$ does not span a CES. Therefore, the theorem follows.

\section{VI. P\lowercase{roof of} P\lowercase{roposition} 1}

\textbf{\textit{Proposition:}}
The orthogonal complement of the set $\mathcal{U}$ constitutes a GES in $2\bigotimes2\bigotimes2$ Hilbert space.

Proof:
The states of the set $\mathcal{U}$ are given by
\begin{equation}
\mathcal{U}:\equiv
\begin{array}{l}
 \ket{\phi\down{0}{\ssty 0}^{\ssty-}}\move{2}= \ket{0}(\ket{00}-2\ket{10}+\ket{11}),\\[2pt]
 \ket{\phi\down{0}{\ssty 1}^{\ssty-}}\move{2}= (\ket{01}+2\ket{00}-3\ket{10})\ket{0},\\[2pt]
  \ket{\psi\down{1}{\ssty 0}^{\ssty+}}= \ket{1}\ket{01+10},\\[2pt]
  \ket{\psi\down{1}{\ssty 1}^{\ssty+}}= \ket{00+11}\ket{1},\\[2pt]
\ket{\kappa\,}\move{7}= \ket{0+1}\ket{00+10}+\ket{0-1}\ket{01+11}
 \end{array}
\label{UBB}
\end{equation}

So far we can remember the entangled stopper $\ket{\kappa}$ excludes the following states from the cross-L structure:
\begin{equation}
\mathcal{U}^\prime:\equiv
\begin{array}{l}
 \ket{\phi\down{0}{\ssty 0}^{\ssty+}}\move{2}= \ket{0}(\ket{00}-2\ket{10}-5\ket{11}),\\[2pt]
 \ket{\phi\down{0}{\ssty 1}^{\ssty+}}\move{2}= (\ket{01}+2\ket{00}+\frac{5}{3}\ket{10})\ket{0},\\[2pt]
  \ket{\psi\down{1}{\ssty 0}^{\ssty-}}= \ket{1}\ket{01-10},\\[2pt]
  \ket{\psi\down{1}{\ssty 1}^{\ssty-}}= \ket{11-00}\ket{1}
 \end{array}
\end{equation}


Let us define,
\begin{equation}
\Omega=\left\{\left(\move{2}\ketbra{\phi}{\psi\down{1}{\ssty 0}^{\ssty-}}-\ketbra{\psi\down{1}{\ssty 0}^{\ssty-}}{\phi}\move{2}\right)\ket{\kappa\,}, \ket{\phi}\in \mathcal{U}^\prime\setminus\left\{\ket{\psi\down{1}{\ssty 0}^{\ssty-}}\right\}\right\}
\end{equation}

By definition, each state of $\Omega$ is orthogonal to $\ket{\kappa\,}$. Therefore, $\Omega$ spans orthogonal complement of $\mathcal{U}$. The states of the set $\Omega$ can be written as
\begin{equation}
\begin{array}{l}
-2\ket{0}(\ket{00}-2\ket{10}-5\ket{11})+6\ket{1}\ket{01-10},\\[2pt]
-2(\ket{01}+2\ket{00}+\frac{5}{3}\ket{10})\ket{0}-\frac{14}{3}\ket{1}\ket{01-10},\\[2pt]
-2\ket{11-00}\ket{1}+2\ket{1}\ket{01-10}
 \end{array}
\label{UBBcomplement}
\end{equation}

Equivalently, $\Omega$ can be expressed as
\begin{equation}
	\Omega
	:\equiv
	\begin{array}{l}
		\ket{\psi\down{1}{0}}= \ket{0}(\ket{00}-2\ket{10}-5\ket{11})-3\ket{1}\ket{01-10},\\[2pt]
		\ket{\psi\down{1}{1}}= (6\ket{00}+3\ket{01}+5\ket{10})\ket{0}+7\ket{1}\ket{01-10},\\[2pt]
		\ket{\psi\down{1}{2}}= \ket{00-11}\ket{1}+\ket{1}\ket{01-10}
	\end{array}
	\label{UBBcomplement}
\end{equation}

Case 1: 
Consider the system in $A|BC$ partition. In the joint subsyste m $BC$, we introduce the relabeled basis $\{\ket{0}\equiv\ket{00},\ket{1}\equiv\ket{01},\ket{2}\equiv\ket{10},\ket{3}\equiv\ket{11}\}_{\move{0}\ssty BC}$. The bipartite representation of $\Omega$ thus be
\begin{equation}
\begin{array}{l}
 \ket{\psi\down{1}{0}}= \ket{0}(\ket{0}-2\ket{2}-5\ket{3})-3\ket{1}\ket{1-2},\\[2pt]
 \ket{\psi\down{1}{1}}= 6\ket{00}+3\ket{02}+5\ket{10}+7\ket{1}\ket{1-2},\\[2pt]
 \ket{\psi\down{1}{2}}= \ket{01-13}+\ket{1}\ket{1-2}
 \end{array}
\end{equation}

\begin{table}[h]
    \centering
    \caption{$a_{kl}^{(i)}\left(\ket{\psi\down{1}{i}}_{\move{-1.5}\ssty A|BC}=\sum_{kl} a_{kl}^{(i)}\ket{kl}_{\move{-1.5}\ssty A|BC}\in\Omega\right)$}
    \begin{tabular}{c|c|cccc}
        \toprule
        $i$ & $k \setminus l$ & 0 & 1 & 2 & 3 \\
        \midrule
 0 & 0 & \move{3}1 & \move{3}0 & -2 & -5\\
 & 1 & \move{3}0 & -3 & \move{3}3 & \move{3}0\\
\midrule
 1 & 0 & \move{3}6 & \move{3}0 & \move{3}3 & \move{3}0\\
 & 1 & \move{3}5 & \move{3}7 & -7 & \move{3}0\\
\midrule
 2 & 0 & \move{3}0 & \move{3}1 & \move{3}0 & \move{3}0\\
 & 1 & \move{3}0 & \move{3}1 & -1 & -1\\
\bottomrule
    \end{tabular}
    \label{table_1}
\end{table}

The product forming matrices of $\Omega$ in $A|BC$ partition are
\begin{equation}
\zoom{c}{\begin{array}{ll}
P_0=\Lambda_{\ssty 01|01}^{\ssty    A|BC}=
\begin{bmatrix}
-3&\move{8}7&1\\
-18&\move{8}42&6\\
\move{8}0&-5&0
\end{bmatrix},
&P_1=\Lambda_{\ssty 01|02}^{\ssty    A|BC}=
\begin{bmatrix}
\move{3}3 & \move{8}3 & -1 \\
\move{3}18 & -57 & -6 \\
\move{3}0 & \move{8}0 & \move{8}0
\end{bmatrix},\\[15pt]
P_2=\Lambda_{\ssty 01|03}^{\ssty    A|BC} =
\begin{bmatrix}
\move{8}0 &\move{5} 25 & -1 \\
\move{8}0 & \move{5}0 & -6 \\
\move{8}0 & \move{5}0 & \move{8}0
\end{bmatrix},
&P_3=\Lambda_{\ssty 01|12}^{\ssty    A|BC}=
\begin{bmatrix}
-6 & \move{8}14 & \move{8}2\\
\move{8}9 & -21 & -3\\
\move{8}3 & -7 & -1
\end{bmatrix},\\[15pt]
P_4=\Lambda_{\ssty 01|13}^{\ssty    A|BC}=
\begin{bmatrix}
-15 & 35 & \move{8}5\\
\move{8}0 & 0 & \move{8}0\\
\move{8}0 & 0 & -1
\end{bmatrix},
&P_5=\Lambda_{\ssty 01|23}^{\ssty    A|BC}=
\begin{bmatrix}
\move{3}15 & -35 & -3\\
\move{3}0 & \move{8}0 & -3\\
\move{3}0 & \move{8}0 & \move{8}0
\end{bmatrix}\\[20pt]
\end{array}
\label{product_forming_matrices_A|BC}}
\end{equation}

These matrices are obtained directly from the $a$-coefficients listed in Table \ref{table_1}.

For $X=(x_0,\,x_1,\,x_2)^t$, the Eqs. (\ref{2x2_block}) can then be expressed as:
\begin{equation}
\begin{array}{l}
- 3 x_{0}^{2} - 11 x_{0} x_{1} + x_{0} x_{2} + 42 x_{1}^{2} + x_{1} x_{2} = 0,\\[2pt]
\move{7}3 x_{0}^{2} + 21 x_{0} x_{1} - x_{0} x_{2} - 57 x_{1}^{2} - 6 x_{1} x_{2} = 0,\\[2pt]
\move{7}25 x_{0} x_{1} - x_{0} x_{2} - 6 x_{1} x_{2} = 0,\\[2pt]
- 6 x_{0}^{2} + 23 x_{0} x_{1} + 5 x_{0} x_{2} - 21 x_{1}^{2} - 10 x_{1} x_{2} - x_{2}^{2} = 0,\\[2pt]
- 15 x_{0}^{2} + 35 x_{0} x_{1} + 5 x_{0} x_{2} - x_{2}^{2} = 0,\\[2pt]
\move{7}15 x_{0}^{2} - 35 x_{0} x_{1} - 3 x_{0} x_{2} - 3 x_{1} x_{2} = 0
\end{array}
\end{equation}

Using Gröbner basis formalism, we get the simplified form:
\begin{equation}
\begin{array}{ll}
&x_0^{2}=x_1^{2}=x_2^{2}=0\\
\Rightarrow& x_0=x_1=x_2=0$$
\end{array}
\end{equation}

Therefore, by Lemma \ref{product_condition}, the orthogonal complement of $\mathcal{U}$ spans a CES in $A|BC$ bipartition. 

Case 2:
The bipartite representation of $\Omega$ in $B|CA$ will be
\begin{equation}
	\begin{array}{l}
		\ket{\psi\down{1}{0}}= \ket{00}-2\ket{10}-5\ket{12}-3\ket{03-11},\\[2pt]
		\ket{\psi\down{1}{1}}= 6\ket{00}+3\ket{10}+5\ket{01}+7\ket{03-11},\\[2pt]
		\ket{\psi\down{1}{2}}= \ket{02-13}+\ket{03-11}
	\end{array}
\end{equation}

\begin{table}[h]
    \centering
    \caption{$a_{kl}^{(i)}\left(\ket{\psi\down{1}{i}}_{\move{-1.5}\ssty B|CA}=\sum_{kl} a_{kl}^{(i)}\ket{kl}_{\move{-1.5}\ssty B|CA}\in\Omega\right)$}
    \begin{tabular}{c|c|cccc}
        \toprule
        $i$ & $k \setminus l$ & 0 & 1 & 2 & 3 \\
        \midrule
 0 & 0 & \move{3}1 & \move{3}0 & \move{3}0 & -3\\
 & 1 & -2 & \move{3}3 & -5 & \move{3}0\\
\midrule
 1 & 0 & \move{3}6 & \move{3}5 & \move{3}0 & \move{3}7\\
 & 1 & \move{3}3 & -7 & \move{3}0 & \move{3}0\\
\midrule
 2 & 0 & \move{3}0 & \move{3}0 & \move{3}1 & \move{3}1\\
 & 1 & \move{3}0 & -1 & \move{3}0 & -1\\
\bottomrule
    \end{tabular}
    \label{table_2}
\end{table}

The product forming matrices of $\Omega$ in $B|CA$ partition are
\begin{equation}
\zoom{c}{\begin{array}{ll}
P_0=\Lambda_{\ssty 01|01}^{\ssty    B|CA}=
\begin{bmatrix}
  \move{8}3 & -7 & -1 & \\
  \move{8}28 & -57 & -6 & \\
  \move{8}0 & \move{8}0 & \move{8}0 & \\
\end{bmatrix},
&P_1=\Lambda_{\ssty 01|02}^{\ssty    B|CA}=
\begin{bmatrix}
  -5 & \move{8}0 & \move{8}0 & \\
  -30 & \move{8}0 & \move{8}0 & \\
  \move{8}2 & -3 & \move{8}0 & \\
\end{bmatrix},\\[15pt]
P_2=\Lambda_{\ssty 01|03}^{\ssty    B|CA} =
\begin{bmatrix}
  -6 & \move{8}9 & -1 & \\
  \move{8}14 & -21 & -6 & \\
  \move{8}2 & -3 & \move{8}0 & \\
\end{bmatrix},
&P_3=\Lambda_{\ssty 01|12}^{\ssty    B|CA}=
\begin{bmatrix}
  \move{8}0 & \move{8}0 & \move{8}0 & \\
  -25 & \move{8}0 & \move{8}0 & \\
  -3 & \move{8}7 & \move{8}1 & \\
\end{bmatrix},\\[15pt]
P_4=\Lambda_{\ssty 01|13}^{\ssty    B|CA}=
\begin{bmatrix}
  \move{8}9 & -21 & -3 & \\
  -21 & \move{8}49 & \move{8}2 & \\
  -3 & \move{8}7 & \move{8}1 & \\
\end{bmatrix},
&P_5=\Lambda_{\ssty 01|23}^{\ssty    B|CA}=
\begin{bmatrix}
  -15 & \move{8}0 & \move{8}0 & \\
  \move{8}35 & \move{8}0 & \move{8}0 & \\
  \move{8}5 & \move{8}0 & -1 & \\
\end{bmatrix}\\[20pt]
\end{array}
\label{product_forming_matrices_B|CA}}
\end{equation}

These matrices are obtained directly from the $a$-coefficients listed in Table \ref{table_2}.

For $X=(x_0,\,x_1,\,x_2)^t$, the Eqs. (\ref{2x2_block}) can then be expressed as:
\begin{equation}
\begin{array}{l}
\move{7}3 x_{0}^{2} + 21 x_{0} x_{1} - x_{0} x_{2} - 57 x_{1}^{2} - 6 x_{1} x_{2} = 0,\\[2pt]
- 5 x_{0}^{2} - 30 x_{0} x_{1} + 2 x_{0} x_{2} - 3 x_{1} x_{2} = 0,\\[2pt]
- 6 x_{0}^{2} + 23 x_{0} x_{1} + x_{0} x_{2} - 21 x_{1}^{2} - 9 x_{1} x_{2} = 0,\\[2pt]
- 25 x_{0} x_{1} - 3 x_{0} x_{2} + 7 x_{1} x_{2} + x_{2}^{2} = 0,\\[2pt]
\move{7}9 x_{0}^{2} - 42 x_{0} x_{1} - 6 x_{0} x_{2} + 49 x_{1}^{2} + 9 x_{1} x_{2} + x_{2}^{2} = 0,\\[2pt]
- 15 x_{0}^{2} + 35 x_{0} x_{1} + 5 x_{0} x_{2} - x_{2}^{2} = 0
\end{array}
\end{equation}

Using Gröbner basis formalism, we get the simplified form:
\begin{equation}
\begin{array}{ll}
&x_0^{2}=x_1^{2}=x_2^{2}=0\\
\Rightarrow& x_0=x_1=x_2=0$$
\end{array}
\end{equation}

Therefore, by Lemma \ref{product_condition}, the orthogonal complement of $\mathcal{U}$ spans a CES in $B|CA$ bipartition. 

Case 3:
The bipartite representation of $\Omega$ in $C|AB$ will be
\begin{equation}
	\begin{array}{l}
		\ket{\psi\down{1}{0}}= \ket{00}-2\ket{01}-5\ket{11}-3\ket{12-03},\\[2pt]
		\ket{\psi\down{1}{1}}= \ket{0}(6\ket{0}+3\ket{1}+5\ket{2})+7\ket{12-03},\\[2pt]
		\ket{\psi\down{1}{2}}= \ket{1}\ket{0-3}+\ket{12-03}
	\end{array}
\end{equation}

\begin{table}[h]
    \centering
    \caption{$a_{kl}^{(i)}\left(\ket{\psi\down{1}{i}}_{\move{-1.5}\ssty C|AB}=\sum_{kl} a_{kl}^{(i)}\ket{kl}_{\move{-1.5}\ssty C|AB}\in\Omega\right)$}
    \begin{tabular}{c|c|cccc}
        \toprule
        $i$ & $k \setminus l$ & 0 & 1 & 2 & 3 \\
        \midrule
 0 & 0 & \move{3}1 & -2 & \move{3}0 & \move{3}3\\
 & 1 & \move{3}0 & -5 & -3 & \move{3}0\\
\midrule
 1 & 0 & \move{3}6 & \move{3}3 & \move{3}5 & -7\\
 & 1 & \move{3}0 & \move{3}0 & \move{3}7 & \move{3}0\\
\midrule
 2 & 0 & \move{3}0 & \move{3}0 & \move{3}0 & -1\\
 & 1 & \move{3}1 & \move{3}0 & \move{3}1 & -1\\
\bottomrule
    \end{tabular}
    \label{table_3}
\end{table}

The product forming matrices of $\Omega$ in $C|AB$ partition are
\begin{equation}
\zoom{c}{\begin{array}{ll}
P_0=\Lambda_{\ssty 01|01}^{\ssty    C|AB}=
\begin{bmatrix}
  -5 & \move{8}0 & \move{8}2 & \\
  -30 & \move{8}0 & -3 & \\
  \move{8}0 & \move{8}0 & \move{8}0 & \\
\end{bmatrix},
&P_1=\Lambda_{\ssty 01|02}^{\ssty    C|AB}=
\begin{bmatrix}
  -3 & \move{8}7 & \move{8}1 & \\
  -18 & \move{8}42 & \move{8}1 & \\
  \move{8}0 & \move{8}0 & \move{8}0 & \\
\end{bmatrix},\\[15pt]
P_2=\Lambda_{\ssty 01|03}^{\ssty    C|AB} =
\begin{bmatrix}
  \move{8}0 & \move{8}0 & \move{5}-4 & \\
  \move{8}0 & \move{8}0 & \move{13}1 & \\
  \move{8}0 & \move{8}0 & \move{13}1 & \\
\end{bmatrix},
&P_3=\Lambda_{\ssty 01|12}^{\ssty    C|AB}=
\begin{bmatrix}
  \move{8}6 & -14 & -2 & \\
  \move{8}16 & \move{8}21 & \move{8}3 & \\
  \move{8}0 & \move{8}0 & \move{8}0 & \\
\end{bmatrix},\\[15pt]
P_4=\Lambda_{\ssty 01|13}^{\ssty    C|AB}=
\begin{bmatrix}
  \move{8}15 & \move{8}0 & \move{8}2 & \\
  -35 & \move{8}0 & -3 & \\
  -5 & \move{8}0 & \move{8}0 & \\
\end{bmatrix},
&P_5=\Lambda_{\ssty 01|23}^{\ssty    C|AB}=
\begin{bmatrix}
  \move{8}9 & -21 & -3 & \\
  -21 & \move{8}49 & \move{8}2 & \\
  -3 & \move{8}7 & \move{8}1 & \\
\end{bmatrix}\\[20pt]
\end{array}
\label{product_forming_matrices_C|AB}}
\end{equation}

These matrices are obtained directly from the $a$-coefficients listed in Table \ref{table_3}.

For $X=(x_0,\,x_1,\,x_2)^t$, the Eqs. (\ref{2x2_block}) can then be expressed as:
\begin{equation}
\begin{array}{l}
- 5 x_{0}^{2} - 30 x_{0} x_{1} + 2 x_{0} x_{2} - 3 x_{1} x_{2} = 0,\\[2pt]
- 3 x_{0}^{2} - 11 x_{0} x_{1} + x_{0} x_{2} + 42 x_{1}^{2} + x_{1} x_{2} = 0,\\[2pt]
- 4 x_{0} x_{2} + x_{1} x_{2} + x_{2}^{2} = 0,\\[2pt]
\move{7}6 x_{0}^{2} + 2 x_{0} x_{1} - 2 x_{0} x_{2} + 21 x_{1}^{2} + 3 x_{1} x_{2} = 0,\\[2pt]
\move{7}15 x_{0}^{2} - 35 x_{0} x_{1} - 3 x_{0} x_{2} - 3 x_{1} x_{2} = 0,\\[2pt]
\move{7}9 x_{0}^{2} - 42 x_{0} x_{1} - 6 x_{0} x_{2} + 49 x_{1}^{2} + 9 x_{1} x_{2} + x_{2}^{2} = 0
\end{array}
\end{equation}

Using Gröbner basis formalism, we get the simplified form:
\begin{equation}
\begin{array}{ll}
&x_0^{2}=x_1^{2}=x_2^{2}=0\\
\Rightarrow& x_0=x_1=x_2=0$$
\end{array}
\end{equation}

Therefore, by Lemma \ref{product_condition}, the orthogonal complement of $\mathcal{U}$ spans a CES in $C|AB$ bipartition. 

Neither system of quadratic equations admits a nonzero solution. Consequently, the set $\Omega$ spans an entangled subspace in every bipartition; in other words, the orthogonal complement of $\mathcal{U}$ spans a GES.

\section{VII. P\lowercase{roof of} P\lowercase{roposition} 2}

\textbf{\textit{Lemma:}}
Consider an $n$-dimensional subspace $S_{\ssty AB}$ of a bipartite Hilbert space $\mathbb{C}^{d_{\ssty  A}}\otimes\mathbb{C}^{d_{\ssty B}}.$ If the projector $\mathbb{P}_{\move{-1}\ssty AB}$ on $S_{\ssty  AB}$ satisfies the condition $\rank(\mathbb{P}_{\move{-1}\ssty AB})<\max\{\rank(\mathbb{P}_{\move{-1}\ssty A}),\rank(\mathbb{P}_{\move{-1}\ssty B})\}$, then all the rank-$n$ states supported on $\mathcal{S}_{\ssty AB}$ are $1$-distillable; where $\mathbb{P}_{\move{-1}\ssty A(B)}:=\Tr_{\ssty B(A)}(\mathbb{P}_{\move{-1}\ssty AB})$ \cite{Agrawal2019}.

\textbf{\textit{Proposition:}}
The orthogonal complement of the set  $\mathcal{U}$ is distillable across every bipartition.

Proof:
Let $\mathcal{U}^\perp$ be the orthogonal complement of the set $\mathcal{U}$. Consider an arbitrary projector $\mathbb{P}(n)$ of rank $n$ acting on $\mathcal{U}^\perp$. The following facts hold:
(i) Rank of bimarginal of the convex mixture of $n$ mutually orthogonal vectors cannot be less than that of the minimum number of states in $\Omega$ required to construct them.
(ii) To construct $n$ mutually orthogonal vectors in the subspace $\mathcal{U}^\perp$ at least $n$ states from the set $\Omega$ are required.
(iii) Consider arbitrary $n$ number of states $\{\ket{\eta\down{1}{\ssty i}}\}_{\ssty i=0}^{\ssty n-1}$ from $\Omega$. For any such choices 
$\rank [\Tr_{\ssty\alpha}(\sum_{\ssty i=0}^{\ssty n-1}\ketbra{\eta\down{1}{\ssty i}}{\eta\down{1}{\ssty i}})]\geqslant n+1$, where $\alpha\in\{A,B,C\}$.

Therefore, bimarginals of an arbitrary projector $\mathbb{P}(n)$ have rank at least $n+1$. Consequently, the previous lemma guarantees the distillability of $\mathcal{U}^\perp$ across every bipartition.

\section{VIII. P\lowercase{roof of} P\lowercase{roposition} 3}

\textbf{\textit{Proposition:}}
Let $W \subset \mathcal{H}^{2\bigotimes2\bigotimes2}$ be the subspace spanned by any four states of the set $\mathcal{U}$. Then the pair $\{W, W^\perp\}$ forms a 2-CES splitting of $\mathcal{H}$.
 
Proof:
Let $\Lambda^{\ssty A|BC}$, $\Lambda^{\ssty B|CA}$, $\Lambda^{\ssty C|AB}$ denote the sets of all product forming matrices corresponding to the bipartitions $A|BC$, $B|CA$, and $C|AB$, respectively. Since any product state remains a product across every bipartition, each nonzero solution of the homogeneous quadratic system
\begin{equation}
	X^tPX=0,\quad \forall\; P\in \Lambda^{\ssty A|BC}\cup\Lambda^{\ssty B|CA}\cup\Lambda^{\ssty C|AB}
	\label{product_equations}
\end{equation}
is associated with a product state.

Let $\ket{\phi}$ be any state in $\mathcal{U}$. Define
\begin{equation}
	\mathcal{C}=\mathcal{U}\setminus\{\ket{\phi}\},\quad\mathcal{C}^\prime=\Omega\cup\{\ket{\phi}\},
\end{equation}
and
\begin{equation}
	W=\Span({\sty\mathcal{C}}),\quad W^\prime=\Span({\sty\mathcal{C}^\prime})
\end{equation}

By definition $\mathcal{C}\cup\mathcal{C}^\prime$ spans the entire $2\bigotimes2\bigotimes2$ Hilbert space. Since the states in $\mathcal{U}$ are mutually orthogonal and $\Omega$ spans the orthogonal complement of $\mathcal{U}$, it follows that $\mathcal{C}^\prime$ spans the orthogonal complement of $\mathcal{C}$. Hence,
\begin{equation}
	W^\prime=W^\perp
\end{equation}

It therefore suffices to show that neither $W$ nor $W^\prime$ contains a product state, i.e., that the homogeneous quadratic systems associated with $\mathcal{C}$ and $\mathcal{C}^\prime$ admit no nonzero solution in $X=(x_0,x_1,x_2,x_3)^t$. We verify this by explicitly evaluating the corresponding quadratic polynomials and computing their Gröbner bases using a Python algorithm.

Any product state in $W^\prime$ can be written as
\begin{equation}
a_0\ket{\psi\down{1}{0}}+a_1\ket{\psi\down{1}{0}}+a_2\ket{\psi\down{1}{0}}+a_3\ket{\phi}
\end{equation}

Since $\Omega$ spans an entangled subspace, one must have $a_3\neq0$. Consequently, the state 
\begin{equation}
\frac{a_0}{a_3}\ket{\psi\down{1}{0}}+\frac{a_1}{a_3}\ket{\psi\down{1}{0}}+\frac{a_2}{a_3}\ket{\psi\down{1}{0}}+\ket{\phi}
\end{equation}
is also a product state. This implies that the system above admits a nonzero solution of the form
\begin{equation}
(\frac{a_0}{a_3},\frac{a_1}{a_3},\frac{a_2}{a_3},1)^t
\end{equation}

Equivalently, the nonhomogeneous quadratic system
\begin{equation}
	X\down{1}{3}^tPX\down{1.5}{3}=0,\quad \forall\; P\in \Lambda^{\ssty A|BC}\cup\Lambda^{\ssty B|CA}\cup\Lambda^{\ssty C|AB}
	\label{product_equations_2}
\end{equation}
with $X\down{0}{3}=(x_0,\,x_1,\,x_2,\,1)^t$, would admit a solution.



\textbf{Case 1:}
Let $\ket{\phi}=\ket{\phi\down{0}{\ssty 0}^{\ssty-}}$. Then the states of $\mathcal{C}$ and $\mathcal{C}^\prime$ can be expressed as
\begin{equation}
	\mathcal{C}:\equiv
	\begin{array}{l}
		\ket{\phi\down{0}{\ssty 1}^{\ssty-}}\move{2}= (\ket{01}+2\ket{00}-3\ket{10})\ket{0},\\[2pt]
		\ket{\psi\down{1}{\ssty 0}^{\ssty+}}= \ket{1}\ket{01+10},\\[2pt]
		\ket{\psi\down{1}{\ssty 1}^{\ssty+}}= \ket{00+11}\ket{1},\\[2pt]
		\ket{\kappa\,}\move{7}= \ket{0+1}\ket{00+10}+\ket{0-1}\ket{01+11}
	\end{array}
	\label{UBB}
\end{equation}

\begin{equation}
	\mathcal{C}^\prime:\equiv
	\begin{array}{l}
		\ket{\psi\down{1}{0}}= \ket{0}(\ket{00}-2\ket{10}-5\ket{11})-3\ket{1}\ket{01-10},\\[2pt]
		\ket{\psi\down{1}{1}}= (6\ket{00}+3\ket{01}+5\ket{10})\ket{0}+7\ket{1}\ket{01-10},\\[2pt]
		\ket{\psi\down{1}{2}}= \ket{00-11}\ket{1}+\ket{1}\ket{01-10},\\[2pt]
		\ket{\phi\down{0}{\ssty 0}^{\ssty-}}= \ket{0}(\ket{00}-2\ket{10}+\ket{11})
	\end{array}
\end{equation}

The product forming matrices of $\mathcal{C}$ are
\begin{equation}
	\zoom{c}{\begin{array}{ll}
			P_0\move{3.5}=\Lambda_{\ssty 01|01}^{\ssty    A|BC}=
			\begin{bmatrix}
  \move{8}0 & \move{8}2 & \move{8}0 & -2 & \\
  \move{8}0 & \move{8}0 & \move{8}0 & \move{8}0 & \\
  \move{8}3 & \move{8}0 & \move{8}0 & -1 & \\
  \move{8}3 & \move{8}1 & \move{8}0 & -2 & \\
\end{bmatrix},
			&P_1\move{3.5}=\Lambda_{\ssty 01|02}^{\ssty    A|BC}=
			\begin{bmatrix}
  \move{8}3 & \move{8}2 & \move{8}0 & \move{8}1 & \\
  \move{8}0 & \move{8}0 & \move{8}0 & \move{8}0 & \\
  \move{8}0 & \move{8}0 & \move{8}0 & \move{8}0 & \\
  \move{8}3 & \move{8}1 & \move{8}0 & \move{8}0 & \\
\end{bmatrix},\\[20pt]
			P_2\move{3.5}=\Lambda_{\ssty 01|03}^{\ssty    A|BC} =
			\begin{bmatrix}
  \move{8}0 & \move{8}0 & \move{8}2 & -2 & \\
  \move{8}0 & \move{8}0 & \move{8}0 & \move{8}0 & \\
  \move{8}0 & \move{8}0 & \move{8}0 & \move{8}0 & \\
  \move{8}3 & \move{8}0 & \move{8}1 & -2 & \\
\end{bmatrix},
			&P_3\move{3.5}=\Lambda_{\ssty 01|12}^{\ssty    A|BC}=
			\begin{bmatrix}
  \move{8}0 & -1 & \move{8}0 & \move{8}1 & \\
  \move{8}0 & \move{8}0 & \move{8}0 & \move{8}0 & \\
  \move{8}0 & \move{8}1 & \move{8}0 & \move{8}1 & \\
  \move{8}0 & \move{8}0 & \move{8}0 & \move{8}2 & \\
\end{bmatrix},\\[20pt]
			P_4\move{3.5}=\Lambda_{\ssty 01|13}^{\ssty    A|BC}=
			\begin{bmatrix}
  \move{8}0 & \move{8}0 & \move{8}0 & \move{8}0 & \\
  \move{8}0 & \move{8}0 & \move{8}0 & \move{8}0 & \\
  \move{8}0 & \move{8}0 & \move{8}1 & -1 & \\
  \move{8}0 & -1 & \move{8}1 & \move{8}0 & \\
\end{bmatrix},
			&P_5\move{3.5}=\Lambda_{\ssty 01|23}^{\ssty    A|BC}=
			\begin{bmatrix}
  \move{8}0 & \move{8}0 & \move{8}1 & -1 & \\
  \move{8}0 & \move{8}0 & \move{8}0 & \move{8}0 & \\
  \move{8}0 & \move{8}0 & \move{8}0 & \move{8}0 & \\
  \move{8}0 & -1 & \move{8}1 & -2 & \\
\end{bmatrix}\\[20pt]
			P_6\move{3.5}=\Lambda_{\ssty 01|01}^{\ssty    B|CA}=
			\begin{bmatrix}
  \move{8}3 & \move{8}2 & \move{8}0 & \move{8}5 & \\
  \move{8}0 & \move{8}0 & \move{8}0 & \move{8}0 & \\
  \move{8}0 & \move{8}0 & \move{8}0 & \move{8}0 & \\
  -1 & \move{8}1 & \move{8}0 & \move{8}0 & \\
\end{bmatrix},
			&P_7\move{3.5}=\Lambda_{\ssty 01|02}^{\ssty    B|CA}=
			\begin{bmatrix}
  \move{8}0 & \move{8}0 & \move{8}0 & \move{8}2 & \\
  \move{8}0 & \move{8}0 & \move{8}0 & \move{8}0 & \\
  -1 & \move{8}0 & \move{8}0 & -1 & \\
  -1 & \move{8}0 & \move{8}0 & \move{8}0 & \\
\end{bmatrix},\\[20pt]
			P_8\move{3.5}=\Lambda_{\ssty 01|03}^{\ssty    B|CA} =
			\begin{bmatrix}
  \move{8}0 & \move{8}0 & \move{8}2 & -2 & \\
  -1 & \move{8}0 & \move{8}0 & -1 & \\
  \move{8}0 & \move{8}0 & \move{8}0 & \move{8}0 & \\
  \move{8}1 & \move{8}0 & \move{8}1 & \move{8}0 & \\
\end{bmatrix},
			&P_9\move{3.5}=\Lambda_{\ssty 01|12}^{\ssty    B|CA}=
			\begin{bmatrix}
  \move{8}0 & \move{8}0 & \move{8}0 & -3 & \\
  \move{8}0 & \move{8}0 & \move{8}0 & \move{8}0 & \\
  \move{8}0 & -1 & \move{8}0 & -1 & \\
  \move{8}0 & -1 & \move{8}0 & \move{8}0 & \\
\end{bmatrix},\\[20pt]
			P_{10}=\Lambda_{\ssty 01|13}^{\ssty    B|CA}=
			\begin{bmatrix}
  \move{8}0 & \move{8}0 & -3 & \move{8}3 & \\
  \move{8}0 & -1 & \move{8}0 & -1 & \\
  \move{8}0 & \move{8}0 & \move{8}0 & \move{8}0 & \\
  \move{8}0 & \move{8}1 & \move{8}1 & \move{8}0 & \\
\end{bmatrix},
			&P_{11}=\Lambda_{\ssty 01|23}^{\ssty    B|CA}=
			\begin{bmatrix}
  \move{8}0 & \move{8}0 & \move{8}0 & \move{8}0 & \\
  \move{8}0 & \move{8}0 & \move{8}0 & -1 & \\
  \move{8}0 & \move{8}0 & \move{8}1 & -1 & \\
  \move{8}0 & \move{8}0 & \move{8}1 & \move{8}0 & \\
\end{bmatrix}\\[20pt]
			P_{12}=\Lambda_{\ssty 01|01}^{\ssty    C|AB}=
			\begin{bmatrix}
  \move{8}0 & \move{8}0 & -1 & \move{8}1 & \\
  \move{8}0 & \move{8}0 & \move{8}0 & \move{8}0 & \\
  \move{8}0 & \move{8}0 & \move{8}0 & \move{8}0 & \\
  \move{8}0 & \move{8}0 & -1 & \move{8}0 & \\
\end{bmatrix},
			&P_{13}=\Lambda_{\ssty 01|02}^{\ssty    C|AB}=
			\begin{bmatrix}
  \move{8}0 & \move{8}2 & \move{8}3 & \move{8}1 & \\
  \move{8}0 & \move{8}0 & \move{8}0 & \move{8}0 & \\
  \move{8}0 & \move{8}0 & \move{8}0 & \move{8}0 & \\
  \move{8}0 & \move{8}1 & -1 & -2 & \\
\end{bmatrix},\\[20pt]
			P_{14}=\Lambda_{\ssty 01|03}^{\ssty    C|AB} =
			\begin{bmatrix}
  \move{8}0 & \move{8}0 & \move{8}2 & -2 & \\
  \move{8}0 & \move{8}0 & -1 & -1 & \\
  \move{8}0 & \move{8}0 & \move{8}0 & \move{8}0 & \\
  \move{8}0 & \move{8}0 & \move{8}0 & -2 & \\
\end{bmatrix},
			&P_{15}=\Lambda_{\ssty 01|12}^{\ssty    C|AB}=
			\begin{bmatrix}
  \move{8}0 & \move{8}1 & \move{8}0 & \move{8}2 & \\
  \move{8}0 & \move{8}0 & \move{8}0 & \move{8}0 & \\
  \move{8}0 & \move{8}0 & \move{8}0 & \move{8}0 & \\
  \move{8}0 & \move{8}1 & \move{8}0 & -2 & \\
\end{bmatrix},\\[20pt]
			P_{16}=\Lambda_{\ssty 01|13}^{\ssty    C|AB}=
			\begin{bmatrix}
  \move{8}0 & \move{8}0 & \move{8}1 & -1 & \\
  \move{8}0 & \move{8}0 & \move{8}0 & -1 & \\
  \move{8}0 & \move{8}0 & \move{8}0 & \move{8}0 & \\
  \move{8}0 & \move{8}0 & \move{8}1 & -2 & \\
\end{bmatrix},
			&P_{17}=\Lambda_{\ssty 01|23}^{\ssty    C|AB}=
			\begin{bmatrix}
  \move{8}0 & \move{8}0 & -3 & \move{8}3 & \\
  \move{8}0 & -1 & \move{8}0 & \move{8}1 & \\
  \move{8}0 & \move{8}0 & \move{8}0 & \move{8}0 & \\
  \move{8}0 & -1 & \move{8}1 & \move{8}0 & \\
\end{bmatrix}\\[20pt]
		\end{array}
}
\end{equation}

The quadratic polynomials in (\ref{product_equations}) can then be written as:
\begin{equation}
\begin{array}{l}
\move{7}2 x_{0} x_{1} + 3 x_{0} x_{2} + x_{0} x_{3} + x_{1} x_{3} - x_{2} x_{3} - 2 x_{3}^{2} = 0,\\[2pt]
\move{7}3 x_{0}^{2} + 2 x_{0} x_{1} + 4 x_{0} x_{3} + x_{1} x_{3} = 0,\\[2pt]
\move{7}2 x_{0} x_{2} + x_{0} x_{3} + x_{2} x_{3} - 2 x_{3}^{2} = 0,\\[2pt]
- x_{0} x_{1} + x_{0} x_{3} + x_{1} x_{2} + x_{2} x_{3} + 2 x_{3}^{2} = 0,\\[2pt]
- x_{1} x_{3} + x_{2}^{2} = 0,\\[2pt]
\move{7}x_{0} x_{2} - x_{0} x_{3} - x_{1} x_{3} + x_{2} x_{3} - 2 x_{3}^{2} = 0,\\[2pt]
\move{7}3 x_{0}^{2} + 2 x_{0} x_{1} + 4 x_{0} x_{3} + x_{1} x_{3} = 0,\\[2pt]
- x_{0} x_{2} + x_{0} x_{3} - x_{2} x_{3} = 0,\\[2pt]
- x_{0} x_{1} + 2 x_{0} x_{2} - x_{0} x_{3} - x_{1} x_{3} + x_{2} x_{3} = 0,\\[2pt]
- 3 x_{0} x_{3} - x_{1} x_{2} - x_{1} x_{3} - x_{2} x_{3} = 0,\\[2pt]
- 3 x_{0} x_{2} + 3 x_{0} x_{3} - x_{1}^{2} + x_{2} x_{3} = 0,\\[2pt]
- x_{1} x_{3} + x_{2}^{2} = 0,\\[2pt]
- x_{0} x_{2} + x_{0} x_{3} - x_{2} x_{3} = 0,\\[2pt]
\move{7}2 x_{0} x_{1} + 3 x_{0} x_{2} + x_{0} x_{3} + x_{1} x_{3} - x_{2} x_{3} - 2 x_{3}^{2} = 0,\\[2pt]
\move{7}2 x_{0} x_{2} - 2 x_{0} x_{3} - x_{1} x_{2} - x_{1} x_{3} - 2 x_{3}^{2} = 0,\\[2pt]
\move{7}x_{0} x_{1} + 2 x_{0} x_{3} + x_{1} x_{3} - 2 x_{3}^{2} = 0,\\[2pt]
\move{7}x_{0} x_{2} - x_{0} x_{3} - x_{1} x_{3} + x_{2} x_{3} - 2 x_{3}^{2} = 0,\\[2pt]
- 3 x_{0} x_{2} + 3 x_{0} x_{3} - x_{1}^{2} + x_{2} x_{3} = 0
\end{array}
\end{equation}

Using Gröbner basis formalism, we get the simplified form:
\begin{equation}
\begin{array}{ll}
&x_{0}^{2} + 2 x_{3}^{2}=x_{1}^{2} - 4 x_{3}^{2}=x_{2}^{2} + 2 x_{3}^{2}=x_{3}^{3}=0\\
\Rightarrow& x_0=x_1=x_2=x_3=0
\end{array}
\end{equation}

Therefore, by Lemma \ref{product_condition}, $W$ does not contain any product state in $2\bigotimes2\bigotimes2$ Hilbert space. 

The product forming matrices of $\mathcal{C}^\prime$ are
\begin{equation}
	\zoom{c}{\begin{array}{ll}
			P_0\move{3.5}=\Lambda_{\ssty 01|01}^{\ssty    A|BC}=
			\begin{bmatrix}
  -3 & \move{8}7 & \move{8}1 & \move{8}0 & \\
  -18 & \move{8}42 & \move{8}6 & \move{8}0 & \\
  \move{8}0 & -5 & \move{8}0 & \move{8}0 & \\
  -3 & \move{8}7 & \move{8}1 & \move{8}0 & \\
\end{bmatrix},
			&P_1\move{3.5}=\Lambda_{\ssty 01|02}^{\ssty    A|BC}=
			\begin{bmatrix}
  \move{8}3 & \move{8}3 & -1 & \move{6}0 & \\
  \move{8}18 & -57 & -6 & \move{6}0 & \\
  \move{8}0 & \move{8}0 & \move{8}0 & \move{6}0 & \\
  \move{8}3 & \move{8}3 & -1 & \move{6}0 & \\
\end{bmatrix},\\[20pt]
			P_2\move{3.5}=\Lambda_{\ssty 01|03}^{\ssty    A|BC} =
			\begin{bmatrix}
  \move{10}0 & \move{10}25 & \move{1}-1 & \move{8}0 & \\
  \move{10}0 & \move{10}0 & \move{1}-6 & \move{8}0 & \\
  \move{10}0 & \move{10}0 & \move{9}0 & \move{8}0 & \\
  \move{10}0 & \move{2}-5 & \move{1}-1 & \move{8}0 & \\
\end{bmatrix},
			&P_3\move{3.5}=\Lambda_{\ssty 01|12}^{\ssty    A|BC}=
			\begin{bmatrix}
  \move{2}-6 & \move{8}14 & \move{8}2 & \move{9}0 & \\
  \move{10}9 & -21 & -3 & \move{9}0 & \\
  \move{10}3 & -7 & -1 & \move{9}0 & \\
  \move{2}-6 & \move{8}14 & \move{8}2 & \move{9}0 & \\
\end{bmatrix},\\[20pt]
			P_4\move{3.5}=\Lambda_{\ssty 01|13}^{\ssty    A|BC}=
			\begin{bmatrix}
  -15 & \move{8}35 & \move{8}5 & \move{8}0 & \\
  \move{8}0 & \move{8}0 & \move{8}0 & \move{8}0 & \\
  \move{8}0 & \move{8}0 & -1 & \move{8}0 & \\
  \move{8}3 & -7 & -1 & \move{8}0 & \\
\end{bmatrix},
			&P_5\move{3.5}=\Lambda_{\ssty 01|23}^{\ssty    A|BC}=
			\begin{bmatrix}
  \move{8}15 & -35 & -3 & \move{6}0 & \\
  \move{8}0 & \move{8}0 & -3 & \move{6}0 & \\
  \move{8}0 & \move{8}0 & \move{8}0 & \move{6}0 & \\
  -3 & \move{8}7 & \move{8}3 & \move{6}0 & \\
\end{bmatrix}\\[20pt]
			P_6\move{3.5}=\Lambda_{\ssty 01|01}^{\ssty    B|CA}=
			\begin{bmatrix}
  \move{8}3 & -7 & -1 & \move{3}0 & \\
  \move{8}28 & -57 & -6 & \move{3}10 & \\
  \move{8}0 & \move{8}0 & \move{8}0 & \move{3}0 & \\
  \move{8}3 & -7 & -1 & \move{3}0 & \\
\end{bmatrix},
			&P_7\move{3.5}=\Lambda_{\ssty 01|02}^{\ssty    B|CA}=
			\begin{bmatrix}
  -5 & \move{10}0 & \move{10}0 & \move{8}1 & \\
  -30 & \move{10}0 & \move{10}0 & \move{8}6 & \\
  \move{8}2 & \move{2}-3 & \move{10}0 & \move{8}2 & \\
  -5 & \move{10}0 & \move{10}0 & \move{8}1 & \\
\end{bmatrix},\\[20pt]
			P_8\move{3.5}=\Lambda_{\ssty 01|03}^{\ssty    B|CA} =
			\begin{bmatrix}
  -6 & \move{8}9 & -1 & -6 & \\
  \move{8}14 & -21 & -6 & \move{3}14 & \\
  \move{8}2 & -3 & \move{8}0 & \move{3}2 & \\
  \move{8}0 & \move{8}0 & -1 & \move{3}0 & \\
\end{bmatrix},
			&P_9\move{3.5}=\Lambda_{\ssty 01|12}^{\ssty    B|CA}=
			\begin{bmatrix}
  \move{8}0 & \move{10}0 & \move{10}0 & \move{8}0 & \\
  -25 & \move{10}0 & \move{10}0 & \move{8}5 & \\
  -3 & \move{10}7 & \move{10}1 & \move{8}0 & \\
  \move{8}0 & \move{10}0 & \move{10}0 & \move{8}0 & \\
\end{bmatrix},\\[20pt]
			P_{10}=\Lambda_{\ssty 01|13}^{\ssty    B|CA}=
\begin{bmatrix}
  \move{8}9 & -21 & -3 & \move{8}0 & \\
  -21 & \move{8}49 & \move{8}2 & \move{8}0 & \\
  -3 & \move{8}7 & \move{8}1 & \move{8}0 & \\
  \move{8}0 & \move{8}0 & \move{8}0 & \move{8}0 & \\
\end{bmatrix},
			&P_{11}=\Lambda_{\ssty 01|23}^{\ssty    B|CA}=
			\begin{bmatrix}
  -15 & \move{10}0 & \move{10}0 & \move{8}3 & \\
  \move{8}35 & \move{10}0 & \move{10}0 & -7 & \\
  \move{8}5 & \move{10}0 & \move{2}-1 & -1 & \\
  \move{8}0 & \move{10}0 & \move{10}0 & \move{8}0 & \\
\end{bmatrix}\\[20pt]
			P_{12}=\Lambda_{\ssty 01|01}^{\ssty    C|AB}=
			\begin{bmatrix}
  -5 & \move{11}0 & \move{10}2 & \move{8}1 & \\
  -30 & \move{11}0 & \move{2}-3 & \move{8}6 & \\
  \move{8}0 & \move{11}0 & \move{10}0 & \move{8}0 & \\
  -5 & \move{11}0 & \move{10}2 & \move{8}1 & \\
\end{bmatrix},
			&P_{13}=\Lambda_{\ssty 01|02}^{\ssty    C|AB}=
			\begin{bmatrix}
  -3 & \move{8}7 & \move{8}1 & \move{8}0 & \\
  -18 & \move{8}42 & \move{8}1 & \move{8}0 & \\
  \move{8}0 & \move{8}0 & \move{8}0 & \move{8}0 & \\
  -3 & \move{8}7 & \move{8}1 & \move{8}0 & \\
\end{bmatrix},\\[20pt]
			P_{14}=\Lambda_{\ssty 01|03}^{\ssty    C|AB} =
			\begin{bmatrix}
  \move{10}0 & \move{14}0 & \move{2}-4 & \move{8}0 & \\
  \move{10}0 & \move{14}0 & \move{10}1 & \move{8}0 & \\
  \move{10}0 & \move{14}0 & \move{10}1 & \move{8}0 & \\
  \move{10}0 & \move{14}0 & \move{2}-1 & \move{8}0 & \\
\end{bmatrix},
			&P_{15}=\Lambda_{\ssty 01|12}^{\ssty    C|AB}=
			\begin{bmatrix}
  \move{8}6 & -14 & -2 & \move{8}0 & \\
  \move{8}16 & \move{8}21 & \move{8}3 & -5 & \\
  \move{8}0 & \move{8}0 & \move{8}0 & \move{8}0 & \\
  \move{8}6 & -14 & -2 & \move{8}0 & \\
\end{bmatrix},\\[20pt]
			P_{16}=\Lambda_{\ssty 01|13}^{\ssty    C|AB}=
			\begin{bmatrix}
  \move{8}15 & \move{10}0 & \move{8}2 & \move{2}-3 & \\
  -35 & \move{10}0 & -3 & \move{10}7 & \\
  -5 & \move{10}0 & \move{8}0 & \move{10}1 & \\
  \move{8}0 & \move{10}0 & \move{8}2 & \move{10}0 & \\
\end{bmatrix},
			&P_{17}=\Lambda_{\ssty 01|23}^{\ssty    C|AB}=
			\begin{bmatrix}
  \move{8}9 & -21 & -3 & \move{8}0 & \\
  -21 & \move{8}49 & \move{8}2 & \move{8}0 & \\
  -3 & \move{8}7 & \move{8}1 & \move{8}0 & \\
  \move{8}0 & \move{8}0 & \move{8}0 & \move{8}0 & \\
\end{bmatrix}\\[20pt]
		\end{array}
}
\end{equation}

The quadratic polynomials in (\ref{product_equations_2}) can then be written as:
\begin{equation}
\zoom{c}{
\begin{array}{l}
- 3 x_{0}^{2} - 11 x_{0} x_{1} - 3 x_{0} x_{3} + x_{0} + 42 x_{1}^{2} + 7 x_{1} x_{3} + x_{1} + x_{3} = 0,\\[2pt]
\move{7}3 x_{0}^{2} + 21 x_{0} x_{1} + 3 x_{0} x_{3} - x_{0} - 57 x_{1}^{2} + 3 x_{1} x_{3} - 6 x_{1} - x_{3} = 0,\\[2pt]
\move{7}25 x_{0} x_{1} - x_{0} - 5 x_{1} x_{3} - 6 x_{1} - x_{3} = 0,\\[2pt]
- 6 x_{0}^{2} + 23 x_{0} x_{1} - 6 x_{0} x_{3} + 5 x_{0} - 21 x_{1}^{2} + 14 x_{1} x_{3} - 10 x_{1} + 2 x_{3} - 1 = 0,\\[2pt]
- 15 x_{0}^{2} + 35 x_{0} x_{1} + 3 x_{0} x_{3} + 5 x_{0} - 7 x_{1} x_{3} - x_{3} - 1 = 0,\\[2pt]
\move{7}15 x_{0}^{2} - 35 x_{0} x_{1} - 3 x_{0} x_{3} - 3 x_{0} + 7 x_{1} x_{3} - 3 x_{1} + 3 x_{3} = 0,\\[2pt]
\move{7}3 x_{0}^{2} + 21 x_{0} x_{1} + 3 x_{0} x_{3} - x_{0} - 57 x_{1}^{2} + 3 x_{1} x_{3} - 6 x_{1} - x_{3} = 0,\\[2pt]
- 5 x_{0}^{2} - 30 x_{0} x_{1} - 4 x_{0} x_{3} + 2 x_{0} + 6 x_{1} x_{3} - 3 x_{1} + x_{3}^{2} + 2 x_{3} = 0,\\[2pt]
- 6 x_{0}^{2} + 23 x_{0} x_{1} - 6 x_{0} x_{3} + x_{0} - 21 x_{1}^{2} + 14 x_{1} x_{3} - 9 x_{1} + x_{3} = 0,\\[2pt]
- 25 x_{0} x_{1} - 3 x_{0} + 5 x_{1} x_{3} + 7 x_{1} + 1 = 0,\\[2pt]
\move{7}9 x_{0}^{2} - 42 x_{0} x_{1} - 6 x_{0} + 49 x_{1}^{2} + 9 x_{1} + 1 = 0,\\[2pt]
- 15 x_{0}^{2} + 35 x_{0} x_{1} + 3 x_{0} x_{3} + 5 x_{0} - 7 x_{1} x_{3} - x_{3} - 1 = 0,\\[2pt]
- 5 x_{0}^{2} - 30 x_{0} x_{1} - 4 x_{0} x_{3} + 2 x_{0} + 6 x_{1} x_{3} - 3 x_{1} + x_{3}^{2} + 2 x_{3} = 0,\\[2pt]
- 3 x_{0}^{2} - 11 x_{0} x_{1} - 3 x_{0} x_{3} + x_{0} + 42 x_{1}^{2} + 7 x_{1} x_{3} + x_{1} + x_{3} = 0,\\[2pt]
- 4 x_{0} + x_{1} - x_{3} + 1 = 0,\\[2pt]
\move{7}6 x_{0}^{2} + 2 x_{0} x_{1} + 6 x_{0} x_{3} - 2 x_{0} + 21 x_{1}^{2} - 19 x_{1} x_{3} + 3 x_{1} - 2 x_{3} = 0,\\[2pt]
\move{7}15 x_{0}^{2} - 35 x_{0} x_{1} - 3 x_{0} x_{3} - 3 x_{0} + 7 x_{1} x_{3} - 3 x_{1} + 3 x_{3} = 0,\\[2pt]
\move{7}9 x_{0}^{2} - 42 x_{0} x_{1} - 6 x_{0} + 49 x_{1}^{2} + 9 x_{1} + 1 = 0
\end{array}
}
\end{equation}

The associated Gröbner basis contains the constant polynomial $1$. Hence, by Theorem \ref{Groebner_1}, $W^\prime$ does not contain any product state in $2\bigotimes2\bigotimes2$ Hilbert space.

The remaining cases can be analyzed in the same manner and lead to identical conclusions, thereby completing the proof. For brevity, we omit the corresponding lengthy calculations.

\section{IX. P\lowercase{roof of} L\lowercase{emma} 3}

\textbf{\textit{Lemma:}}
The set $\mathcal{U}$ spans a QCES with product index $6$.

Proof:
The product forming matrices of $\mathcal{U}$ are
\begin{equation}
	\zoom{c}{\begin{array}{ll}
			P_0\move{3.5}=\Lambda_{\ssty 01|01}^{\ssty    A|BC}=
			\begin{bmatrix}
  \move{8}0 & \move{8}0 & \move{8}1 & \move{8}0 & -1 & \\
  \move{8}0 & \move{8}0 & \move{8}2 & \move{8}0 & -2 & \\
  \move{8}0 & \move{8}0 & \move{8}0 & \move{8}0 & \move{8}0 & \\
  \move{8}0 & \move{8}3 & \move{8}0 & \move{8}0 & -1 & \\
  \move{8}0 & \move{8}3 & \move{8}1 & \move{8}0 & -2 & \\
\end{bmatrix},
			&P_1\move{3.5}=\Lambda_{\ssty 01|02}^{\ssty    A|BC}=
			\begin{bmatrix}
  \move{8}0 & -6 & \move{8}1 & \move{8}0 & \move{8}3 & \\
  \move{8}0 & \move{8}3 & \move{8}2 & \move{8}0 & \move{8}1 & \\
  \move{8}0 & \move{8}0 & \move{8}0 & \move{8}0 & \move{8}0 & \\
  \move{8}0 & \move{8}0 & \move{8}0 & \move{8}0 & \move{8}0 & \\
  \move{8}0 & \move{8}3 & \move{8}1 & \move{8}0 & \move{8}0 & \\
\end{bmatrix},\\[25pt]
			P_2\move{3.5}=\Lambda_{\ssty 01|03}^{\ssty    A|BC} =
			\begin{bmatrix}
  \move{8}0 & \move{8}3 & \move{8}0 & \move{8}1 & -2 & \\
  \move{8}0 & \move{8}0 & \move{8}0 & \move{8}2 & -2 & \\
  \move{8}0 & \move{8}0 & \move{8}0 & \move{8}0 & \move{8}0 & \\
  \move{8}0 & \move{8}0 & \move{8}0 & \move{8}0 & \move{8}0 & \\
  \move{8}0 & \move{8}3 & \move{8}0 & \move{8}1 & -2 & \\
\end{bmatrix},
			&P_3\move{3.5}=\Lambda_{\ssty 01|12}^{\ssty    A|BC}=
			\begin{bmatrix}
  \move{8}0 & \move{8}0 & \move{8}2 & \move{8}0 & -2 & \\
  \move{8}0 & \move{8}0 & -1 & \move{8}0 & \move{8}1 & \\
  \move{8}0 & \move{8}0 & \move{8}0 & \move{8}0 & \move{8}0 & \\
  \move{8}0 & \move{8}0 & \move{8}1 & \move{8}0 & \move{8}1 & \\
  \move{8}0 & \move{8}0 & \move{8}0 & \move{8}0 & \move{8}2 & \\
\end{bmatrix},\\[25pt]
			P_4\move{3.5}=\Lambda_{\ssty 01|13}^{\ssty    A|BC}=
			\begin{bmatrix}
  \move{8}0 & \move{8}0 & -1 & \move{8}0 & \move{8}1 & \\
  \move{8}0 & \move{8}0 & \move{8}0 & \move{8}0 & \move{8}0 & \\
  \move{8}0 & \move{8}0 & \move{8}0 & \move{8}0 & \move{8}0 & \\
  \move{8}0 & \move{8}0 & \move{8}0 & \move{8}1 & -1 & \\
  \move{8}0 & \move{8}0 & -1 & \move{8}1 & \move{8}0 & \\
\end{bmatrix},
			&P_5\move{3.5}=\Lambda_{\ssty 01|23}^{\ssty    A|BC}=
			\begin{bmatrix}
  \move{8}0 & \move{8}0 & -1 & -2 & \move{8}1 & \\
  \move{8}0 & \move{8}0 & \move{8}0 & \move{8}1 & -1 & \\
  \move{8}0 & \move{8}0 & \move{8}0 & \move{8}0 & \move{8}0 & \\
  \move{8}0 & \move{8}0 & \move{8}0 & \move{8}0 & \move{8}0 & \\
  \move{8}0 & \move{8}0 & -1 & \move{8}1 & -2 & \\
\end{bmatrix}\\[25pt]
			P_6\move{3.5}=\Lambda_{\ssty 01|01}^{\ssty    B|CA}=
			\begin{bmatrix}
  \move{8}0 & \move{8}0 & \move{8}1 & \move{8}0 & \move{8}1 & \\
  -6 & \move{8}3 & \move{8}2 & \move{8}0 & \move{8}5 & \\
  \move{8}0 & \move{8}0 & \move{8}0 & \move{8}0 & \move{8}0 & \\
  \move{8}0 & \move{8}0 & \move{8}0 & \move{8}0 & \move{8}0 & \\
  \move{8}2 & -1 & \move{8}1 & \move{8}0 & \move{8}0 & \\
\end{bmatrix},
			&P_7\move{3.5}=\Lambda_{\ssty 01|02}^{\ssty    B|CA}=
			\begin{bmatrix}
  \move{8}1 & \move{8}0 & \move{8}0 & \move{8}0 & \move{8}1 & \\
  \move{8}2 & \move{8}0 & \move{8}0 & \move{8}0 & \move{8}2 & \\
  \move{8}0 & \move{8}0 & \move{8}0 & \move{8}0 & \move{8}0 & \\
  \move{8}2 & -1 & \move{8}0 & \move{8}0 & -1 & \\
  \move{8}3 & -1 & \move{8}0 & \move{8}0 & \move{8}0 & \\
\end{bmatrix},\\[25pt]
			P_8\move{3.5}=\Lambda_{\ssty 01|03}^{\ssty    B|CA} =
			\begin{bmatrix}
  \move{8}0 & \move{8}0 & \move{8}0 & \move{8}1 & -1 & \\
  \move{8}0 & \move{8}0 & \move{8}0 & \move{8}2 & -2 & \\
  \move{8}2 & -1 & \move{8}0 & \move{8}0 & -1 & \\
  \move{8}0 & \move{8}0 & \move{8}0 & \move{8}0 & \move{8}0 & \\
  -2 & \move{8}1 & \move{8}0 & \move{8}1 & \move{8}0 & \\
\end{bmatrix},
			&P_9\move{3.5}=\Lambda_{\ssty 01|12}^{\ssty    B|CA}=
			\begin{bmatrix}
  \move{8}0 & \move{8}0 & \move{8}0 & \move{8}0 & \move{8}0 & \\
  -3 & \move{8}0 & \move{8}0 & \move{8}0 & -3 & \\
  \move{8}0 & \move{8}0 & \move{8}0 & \move{8}0 & \move{8}0 & \\
  \move{8}0 & \move{8}0 & -1 & \move{8}0 & -1 & \\
  \move{8}1 & \move{8}0 & -1 & \move{8}0 & \move{8}0 & \\
\end{bmatrix},\\[25pt]
			P_{10}=\Lambda_{\ssty 01|13}^{\ssty    B|CA}=
\begin{bmatrix}
  \move{8}0 & \move{8}0 & \move{8}0 & \move{8}0 & \move{8}0 & \\
  \move{8}0 & \move{8}0 & \move{8}0 & -3 & \move{8}3 & \\
  \move{8}0 & \move{8}0 & -1 & \move{8}0 & -1 & \\
  \move{8}0 & \move{8}0 & \move{8}0 & \move{8}0 & \move{8}0 & \\
  \move{8}0 & \move{8}0 & \move{8}1 & \move{8}1 & \move{8}0 & \\
\end{bmatrix},
			&P_{11}=\Lambda_{\ssty 01|23}^{\ssty    B|CA}=
			\begin{bmatrix}
  \move{8}0 & \move{8}0 & \move{8}0 & \move{8}0 & \move{8}0 & \\
  \move{8}0 & \move{8}0 & \move{8}0 & \move{8}0 & \move{8}0 & \\
  -1 & \move{8}0 & \move{8}0 & \move{8}0 & -1 & \\
  \move{8}0 & \move{8}0 & \move{8}0 & \move{8}1 & -1 & \\
  \move{8}1 & \move{8}0 & \move{8}0 & \move{8}1 & \move{8}0 & \\
\end{bmatrix}\\[25pt]
			P_{12}=\Lambda_{\ssty 01|01}^{\ssty    C|AB}=
			\begin{bmatrix}
  \move{8}1 & \move{8}0 & \move{8}0 & \move{8}2 & \move{8}3 & \\
  \move{8}2 & \move{8}0 & \move{8}0 & -1 & \move{8}1 & \\
  \move{8}0 & \move{8}0 & \move{8}0 & \move{8}0 & \move{8}0 & \\
  \move{8}0 & \move{8}0 & \move{8}0 & \move{8}0 & \move{8}0 & \\
  \move{8}1 & \move{8}0 & \move{8}0 & -1 & \move{8}0 & \\
\end{bmatrix},
			&P_{13}=\Lambda_{\ssty 01|02}^{\ssty    C|AB}=
			\begin{bmatrix}
  \move{8}0 & \move{8}0 & \move{8}1 & \move{8}0 & -1 & \\
  \move{8}0 & \move{8}0 & \move{8}2 & \move{8}3 & \move{8}1 & \\
  \move{8}0 & \move{8}0 & \move{8}0 & \move{8}0 & \move{8}0 & \\
  \move{8}0 & \move{8}0 & \move{8}0 & \move{8}0 & \move{8}0 & \\
  \move{8}0 & \move{8}0 & \move{8}1 & -1 & -2 & \\
\end{bmatrix},\\[25pt]
			P_{14}=\Lambda_{\ssty 01|03}^{\ssty    C|AB} =
			\begin{bmatrix}
  \move{8}0 & \move{8}0 & \move{8}0 & \move{8}1 & -1 & \\
  \move{8}0 & \move{8}0 & \move{8}0 & \move{8}2 & -2 & \\
  \move{8}0 & \move{8}0 & \move{8}0 & -1 & -1 & \\
  \move{8}0 & \move{8}0 & \move{8}0 & \move{8}0 & \move{8}0 & \\
  \move{8}0 & \move{8}0 & \move{8}0 & \move{8}0 & -2 & \\
\end{bmatrix},
			&P_{15}=\Lambda_{\ssty 01|12}^{\ssty    C|AB}=
			\begin{bmatrix}
  \move{8}0 & \move{8}0 & -2 & \move{8}0 & \move{8}2 & \\
  \move{8}3 & \move{8}0 & \move{8}1 & \move{8}0 & \move{8}2 & \\
  \move{8}0 & \move{8}0 & \move{8}0 & \move{8}0 & \move{8}0 & \\
  \move{8}0 & \move{8}0 & \move{8}0 & \move{8}0 & \move{8}0 & \\
  -1 & \move{8}0 & \move{8}1 & \move{8}0 & -2 & \\
\end{bmatrix},\\[25pt]
			P_{16}=\Lambda_{\ssty 01|13}^{\ssty    C|AB}=
			\begin{bmatrix}
  \move{8}0 & \move{8}0 & \move{8}0 & -2 & \move{8}2 & \\
  \move{8}0 & \move{8}0 & \move{8}0 & \move{8}1 & -1 & \\
  -1 & \move{8}0 & \move{8}0 & \move{8}0 & -1 & \\
  \move{8}0 & \move{8}0 & \move{8}0 & \move{8}0 & \move{8}0 & \\
  -1 & \move{8}0 & \move{8}0 & \move{8}1 & -2 & \\
\end{bmatrix},
			&P_{17}=\Lambda_{\ssty 01|23}^{\ssty    C|AB}=
			\begin{bmatrix}
  \move{8}0 & \move{8}0 & \move{8}0 & \move{8}0 & \move{8}0 & \\
  \move{8}0 & \move{8}0 & \move{8}0 & -3 & \move{8}3 & \\
  \move{8}0 & \move{8}0 & -1 & \move{8}0 & \move{8}1 & \\
  \move{8}0 & \move{8}0 & \move{8}0 & \move{8}0 & \move{8}0 & \\
  \move{8}0 & \move{8}0 & -1 & \move{8}1 & \move{8}0 & \\
\end{bmatrix}\\[25pt]
		\end{array}
}
\end{equation}

Since every proper subset of $\mathcal{U}$ spans a CES, a state of the form
\begin{equation}
 a_0\ket{\phi\down{0}{\ssty 0}^{\ssty-}}+a_1\ket{\phi\down{0}{\ssty 1}^{\ssty-}}+a_2\ket{\psi\down{1}{\ssty 0}^{\ssty+}}+a_3\ket{\psi\down{1}{\ssty 1}^{\ssty+}}+a_4\ket{\kappa\,}
\end{equation}
can be a product state only if each $a_i\neq0$, for all $i\in[5]$. Accordingly, for any $k\in[5]$, the vector
\begin{equation}
\left(\frac{a_{\ssty 0}}{a\down{1}{\ssty k}},\frac{a_{\ssty 1}}{a\down{1}{\ssty k}},\frac{a_{\ssty 2}}{a\down{1}{\ssty k}},\frac{a_{\ssty 3}}{a\down{1}{\ssty k}},\frac{a_{\ssty 4}}{a\down{1}{\ssty k}}\right)^{t}
\end{equation} 
provides a nonzero solution of the system (\ref{product_equations}) in $X=(x_0,x_1,x_2,x_3,x_4)^t$. Equivalently, it appears as a solution of the nonhomogeneous quadratic system
\begin{equation}
	X\down{1}{k}^tPX\down{2}{k}=0,\quad \forall\; P\in \Lambda^{\ssty A|BC}\cup\Lambda^{\ssty B|CA}\cup\Lambda^{\ssty C|AB}
	\label{product_equations_3}
\end{equation}
where $X\down{1}{k}\in \mathbb{C}^n$ denotes the perturbed vector obtained by replacing $k$-th entry of $X$ with $1$.

Without loss of generality, consider $k=4$. The corresponding quadratic polynomial in the purterbed vector $X_4=(x_{\ssty0},x_{\ssty1},x_{\ssty2},x_{\ssty3},1)$ can be written as:
\begin{equation}
\zoom{c}{
\begin{array}{l}
\move{7}x\down{0}{\ssty0} x\down{0}{\ssty2} - x\down{0}{\ssty0} + 2 x\down{0}{\ssty1} x\down{0}{\ssty2} + 3 x\down{0}{\ssty1} x\down{0}{\ssty3} + x\down{0}{\ssty1} + x\down{0}{\ssty2} - x\down{0}{\ssty3} - 2 = 0,\\[2pt]
- 6 x\down{0}{\ssty0} x\down{0}{\ssty1} + x\down{0}{\ssty0} x\down{0}{\ssty2} + 3 x\down{0}{\ssty0} + 3 x\down{0}{\ssty1}\up{0}{\ssty2} + 2 x\down{0}{\ssty1} x\down{0}{\ssty2} + 4 x\down{0}{\ssty1} + x\down{0}{\ssty2} = 0,\\[2pt]
\move{7}3 x\down{0}{\ssty0} x\down{0}{\ssty1} + x\down{0}{\ssty0} x\down{0}{\ssty3} - 2 x\down{0}{\ssty0} + 2 x\down{0}{\ssty1} x\down{0}{\ssty3} + x\down{0}{\ssty1} + x\down{0}{\ssty3} - 2 = 0,\\[2pt]
\move{7}2 x\down{0}{\ssty0} x\down{0}{\ssty2} - 2 x\down{0}{\ssty0} - x\down{0}{\ssty1} x\down{0}{\ssty2} + x\down{0}{\ssty1} + x\down{0}{\ssty2} x\down{0}{\ssty3} + x\down{0}{\ssty3} + 2 = 0,\\[2pt]
- x\down{0}{\ssty0} x\down{0}{\ssty2} + x\down{0}{\ssty0} - x\down{0}{\ssty2} + x\down{0}{\ssty3}\up{0}{\ssty2} = 0,\\[2pt]
- x\down{0}{\ssty0} x\down{0}{\ssty2} - 2 x\down{0}{\ssty0} x\down{0}{\ssty3} + x\down{0}{\ssty0} + x\down{0}{\ssty1} x\down{0}{\ssty3} - x\down{0}{\ssty1} - x\down{0}{\ssty2} + x\down{0}{\ssty3} - 2 = 0,\\[2pt]
- 6 x\down{0}{\ssty0} x\down{0}{\ssty1} + x\down{0}{\ssty0} x\down{0}{\ssty2} + 3 x\down{0}{\ssty0} + 3 x\down{0}{\ssty1}\up{0}{\ssty2} + 2 x\down{0}{\ssty1} x\down{0}{\ssty2} + 4 x\down{0}{\ssty1} + x\down{0}{\ssty2} = 0,\\[2pt]
\move{7}x\down{0}{\ssty0}\up{0}{\ssty2} + 2 x\down{0}{\ssty0} x\down{0}{\ssty1} + 2 x\down{0}{\ssty0} x\down{0}{\ssty3} + 4 x\down{0}{\ssty0} - x\down{0}{\ssty1} x\down{0}{\ssty3} + x\down{0}{\ssty1} - x\down{0}{\ssty3} = 0,\\[2pt]
\move{7}2 x\down{0}{\ssty0} x\down{0}{\ssty2} + x\down{0}{\ssty0} x\down{0}{\ssty3} - 3 x\down{0}{\ssty0} - x\down{0}{\ssty1} x\down{0}{\ssty2} + 2 x\down{0}{\ssty1} x\down{0}{\ssty3} - x\down{0}{\ssty1} - x\down{0}{\ssty2} + x\down{0}{\ssty3} = 0,\\[2pt]
- 3 x\down{0}{\ssty0} x\down{0}{\ssty1} + x\down{0}{\ssty0} - 3 x\down{0}{\ssty1} - x\down{0}{\ssty2} x\down{0}{\ssty3} - x\down{0}{\ssty2} - x\down{0}{\ssty3} = 0,\\[2pt]
- 3 x\down{0}{\ssty1} x\down{0}{\ssty3} + 3 x\down{0}{\ssty1} - x\down{0}{\ssty2}\up{0}{\ssty2} + x\down{0}{\ssty3} = 0,\\[2pt]
- x\down{0}{\ssty0} x\down{0}{\ssty2} + x\down{0}{\ssty0} - x\down{0}{\ssty2} + x\down{0}{\ssty3}\up{0}{\ssty2} = 0,\\[2pt]
\move{7}x\down{0}{\ssty0}\up{0}{\ssty2} + 2 x\down{0}{\ssty0} x\down{0}{\ssty1} + 2 x\down{0}{\ssty0} x\down{0}{\ssty3} + 4 x\down{0}{\ssty0} - x\down{0}{\ssty1} x\down{0}{\ssty3} + x\down{0}{\ssty1} - x\down{0}{\ssty3} = 0,\\[2pt]
\move{7}x\down{0}{\ssty0} x\down{0}{\ssty2} - x\down{0}{\ssty0} + 2 x\down{0}{\ssty1} x\down{0}{\ssty2} + 3 x\down{0}{\ssty1} x\down{0}{\ssty3} + x\down{0}{\ssty1} + x\down{0}{\ssty2} - x\down{0}{\ssty3} - 2 = 0,\\[2pt]
\move{7}x\down{0}{\ssty0} x\down{0}{\ssty3} - x\down{0}{\ssty0} + 2 x\down{0}{\ssty1} x\down{0}{\ssty3} - 2 x\down{0}{\ssty1} - x\down{0}{\ssty2} x\down{0}{\ssty3} - x\down{0}{\ssty2} - 2 = 0,\\[2pt]
\move{7}3 x\down{0}{\ssty0} x\down{0}{\ssty1} - 2 x\down{0}{\ssty0} x\down{0}{\ssty2} + x\down{0}{\ssty0} + x\down{0}{\ssty1} x\down{0}{\ssty2} + 2 x\down{0}{\ssty1} + x\down{0}{\ssty2} - 2 = 0,\\[2pt]
- x\down{0}{\ssty0} x\down{0}{\ssty2} - 2 x\down{0}{\ssty0} x\down{0}{\ssty3} + x\down{0}{\ssty0} + x\down{0}{\ssty1} x\down{0}{\ssty3} - x\down{0}{\ssty1} - x\down{0}{\ssty2} + x\down{0}{\ssty3} - 2 = 0,\\[2pt]
- 3 x\down{0}{\ssty1} x\down{0}{\ssty3} + 3 x\down{0}{\ssty1} - x\down{0}{\ssty2}\up{0}{\ssty2} + x\down{0}{\ssty3} = 0
\end{array}
\label{QCES_eqs}
}
\end{equation}

Using Gröbner basis formalism, we get the simplified form:
\begin{equation}
\begin{array}{l}
x_0 = \frac{1}{108}( 115 x_{3}^{5} - 409 x_{3}^{4} + 706 x_{3}^{3} - 1246 x_{3}^{2} + 931 x_{3} - 733), \\[4pt]
x_1 = \frac{1}{54}( 5 x_{3}^{5} - 8 x_{3}^{4} - 28 x_{3}^{3} + 46 x_{3}^{2} - 91 x_{3} + 46), \\[4pt]
x_2 = \frac{1}{36}(-5 x_{3}^{5} - 7 x_{3}^{4} + 22 x_{3}^{3} + 26 x_{3}^{2} + 7 x_{3} + 41),
\label{x_eq}
\end{array}
\end{equation}
and,
\begin{equation}
		5 x_{3}^{6} - 18 x_{3}^{5} + 33 x_{3}^{4} - 60 x_{3}^{3} + 51 x_{3}^{2} - 42 x_{3} + 7 = 0
		\label{x_3_eq}
\end{equation}

Let $x\down{0}{\ssty3}\up{0}{\ssty(k)}$ denote  the $k$-th root of Eq. (\ref{x_3_eq}) and 
\begin{equation}
x\down{0}{\ssty0}\up{0}{\ssty(k)}\ket{\phi\down{0}{\ssty 0}^{\ssty-}}+x\down{0}{\ssty1}\up{0}{\ssty(k)}\ket{\phi\down{0}{\ssty 1}^{\ssty-}}+x\down{0}{\ssty2}\up{0}{\ssty(k)}\ket{\psi\down{1}{\ssty 0}^{\ssty+}}+x\down{0}{\ssty3}\up{0}{\ssty(k)}\ket{\psi\down{1}{\ssty 1}^{\ssty+}}+\ket{\kappa\,}
\label{k_product}
\end{equation}
be the corresponding product state. As Eq. (\ref{x_3_eq}) admits exactly six roots, each root gives rise to a distinct product state of the above form. Hence, the span of $\mathcal{U}$ contains precisely six product states and therefore forms a QCES with product index $6$. This completes the proof.

\begin{figure}[h]
	\centering
	\includegraphics[scale=0.43]{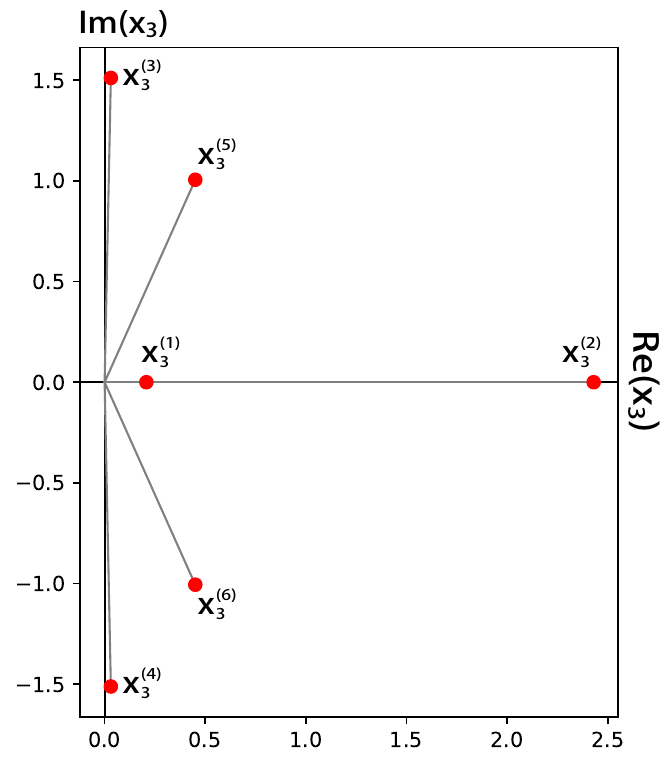}
	\caption{Roots of Eq.(\ref{x_3_eq}) in complex plane.}
	\label{root}
\end{figure}

\begin{table}[h]
    \centering
    \caption{Roots of Eq. (\ref{x_3_eq}) up to six decimal places.}
    \begin{tabular}{c|c}
        \toprule
 $x\down{0}{\ssty3}\up{0}{\ssty(1)}$ & $0.207481$\\
\midrule
 $x\down{0}{\ssty3}\up{0}{\ssty(2)}$ & $2.429704$\\
\midrule
$x\down{0}{\ssty3}\up{0}{\ssty(3)}$ & $0.030984\:+\:\mathrm{i}\:1.511701$ \\
\midrule
$x\down{0}{\ssty3}\up{0}{\ssty(4)}$ & $0.030984\:-\:\mathrm{i}\:1.511701$ \\
\midrule
$x\down{0}{\ssty3}\up{0}{\ssty(5)}$ & $0.450424\:+\:\mathrm{i}\:1.005911$ \\
\midrule
$x\down{0}{\ssty3}\up{0}{\ssty(6)}$ & $0.450424\:-\:\mathrm{i}\:1.005911$ \\
\bottomrule
    \end{tabular}
    \label{table_4}
\end{table}

\begin{table}[h]
    
    \caption{The $k$-th solution of Eqs. (\ref{x_eq}) corresponding to each $x\down{0}{\ssty 3}\up{0}{\ssty(k)}$ (up to six decimal places).}
    \zoom{c}{
    \begin{tabular}{c|c|c|c}
        \toprule
        $k$ & $x\down{0}{\ssty0}\up{0}{\ssty(k)}$ & $x\down{0}{\ssty1}\up{0}{\ssty(k)}$ & $x\down{0}{\ssty2}\up{0}{\ssty(k)}$  \\
        \midrule
 1 & $-5.443347$ & $\move{7}0.534009$ & $\move{7}1.215367$\\
\midrule
 2 & $-2.001367$ & $-2.973833$ & $-3.896768$\\
\midrule
3 & $-0.421030\:-\:\mathrm{i}\:0.612019$ & $-1.732983\:+\:\mathrm{i}\:0.112631$ & $-1.759799\:-\:\mathrm{i}\:2.755537$  \\
\midrule
4 & $-0.421030\:+\:\mathrm{i}\:0.612019$ & $-1.732983\:-\:\mathrm{i}\:0.112631$ & $-1.759799\:+\:\mathrm{i}\:2.755537$ \\
\midrule
5 & $\move{7}0.210054\:+\:\mathrm{i}\:0.235970$ & $\move{7}0.219562\:-\:\mathrm{i}\:0.572381$ & $-0.299500\:+\:\mathrm{i}\:1.002283$ \\
\midrule
6 & \move{7}$0.210054\:-\:\mathrm{i}\:0.235970$ & $\move{7}0.219562\:+\:\mathrm{i}\:0.572381$ & $-0.299500\:-\:\mathrm{i}\:1.002283$ \\
\bottomrule
    \end{tabular}
    }
   \label{table_5}
\end{table}

\section{X. P\lowercase{roof of} P\lowercase{roposition} 4}

\textbf{\textit{Proposition:}}
The orthogonal complement of the subspace spanned by $\mathcal{U}$ is stable even in the extended Hilbert space.

Proof:
Let $\ket{\eta\down{1}{\ssty k}}$ be a product states of the form given in (\ref{k_product}), corresponding to the $k$-th root of Eq. (\ref{x_3_eq}). We aim to show that  $\ket{\eta\down{1}{\ssty k}}+x\ket{\phi}$ is product state if and only if $x=0$, for all $k=1,\ldots,6$ and for every $\ket{\phi}\in \mathcal{U}^{\perp}$.

Any state in $\mathcal{U}^{\perp}$ can be expressed as the superposition of the states in the set $\Omega$ defined in (\ref{UBBcomplement}). Let
\begin{equation}
\begin{array}{r}
x\down{0}{\ssty0}\up{0}{\ssty(k)}\ket{\phi\down{0}{\ssty 0}^{\ssty-}}+x\down{0}{\ssty1}\up{0}{\ssty(k)}\ket{\phi\down{0}{\ssty 1}^{\ssty-}}+x\down{0}{\ssty2}\up{0}{\ssty(k)}\ket{\psi\down{1}{\ssty 0}^{\ssty+}}+x\down{0}{\ssty3}\up{0}{\ssty(k)}\ket{\psi\down{1}{\ssty 1}^{\ssty+}}\\
+\ket{\kappa\,}+x\down{0}{\ssty5}\ket{\psi\down{1}{0}}+x\down{0}{\ssty6}\ket{\psi\down{1}{1}}+x\down{0}{\ssty7}\ket{\psi\down{1}{2}}
\end{array}
\label{k_product_2}
\end{equation}
be a product state for some $x_5,x_6,x_7\in\mathbb{C}$. 

Consider the set
\begin{equation}
\{\ket{\phi\down{0}{\ssty 0}^{\ssty-}},\ket{\phi\down{0}{\ssty 1}^{\ssty-}},\ket{\psi\down{1}{\ssty 0}^{\ssty+}},\ket{\psi\down{1}{\ssty 1}^{\ssty+}},\ket{\kappa\,},\ket{\psi\down{1}{0}},\ket{\psi\down{1}{1}},\ket{\psi\down{1}{2}}\}
\end{equation} 
and the sets of all product forming matrices $\Lambda^{\ssty A|BC}$, $\Lambda^{\ssty B|CA}$, and $\Lambda^{\ssty C|AB}$ corresponding to the bipartitions $A|BC$, $B|CA$, and $C|AB$, respectively.

Consider the variable vector
\begin{equation}
X=(x_{\ssty0},x_{\ssty1},x_{\ssty2},x_{\ssty3},x_{\ssty4},x_{\ssty5},x_{\ssty6},x_{\ssty7})^{t}
\end{equation} 
and the corresponding perturbed vector 
\begin{equation}
X\down{2}{4}=(x_{\ssty0},x_{\ssty1},x_{\ssty2},x_{\ssty3},1,x_{\ssty5},x_{\ssty6},x_{\ssty7})^{t}
\end{equation} 

Therefore, 
\begin{equation}
\begin{array}{r}
x\down{0}{\ssty0}\ket{\phi\down{0}{\ssty 0}^{\ssty-}}+x\down{0}{\ssty1}\ket{\phi\down{0}{\ssty 1}^{\ssty-}}+x\down{0}{\ssty2}\ket{\psi\down{1}{\ssty 0}^{\ssty+}}+x\down{0}{\ssty3}\ket{\psi\down{1}{\ssty 1}^{\ssty+}}\\
+\ket{\kappa\,}+x\down{0}{\ssty5}\ket{\psi\down{1}{0}}+x\down{0}{\ssty6}\ket{\psi\down{1}{1}}+x\down{0}{\ssty7}\ket{\psi\down{1}{2}}
\end{array}
\label{k_product_3}
\end{equation}
is the product state where 
\begin{equation}
\begin{array}{l}
5 x_{3}^{6} - 18 x_{3}^{5} + 33 x_{3}^{4} - 60 x_{3}^{3} + 51 x_{3}^{2} - 42 x_{3} + 7 = 0\\[4pt]
x_0 = \frac{1}{108}( 115 x_{3}^{5} - 409 x_{3}^{4} + 706 x_{3}^{3} - 1246 x_{3}^{2} + 931 x_{3} - 733), \\[4pt]
x_1 = \frac{1}{54}( 5 x_{3}^{5} - 8 x_{3}^{4} - 28 x_{3}^{3} + 46 x_{3}^{2} - 91 x_{3} + 46), \\[4pt]
x_2 = \frac{1}{36}(-5 x_{3}^{5} - 7 x_{3}^{4} + 22 x_{3}^{3} + 26 x_{3}^{2} + 7 x_{3} + 41),

\label{x_eqs}
\end{array}
\end{equation}
and
\begin{equation}
	X\down{1}{4}^tPX\down{2}{4}=0,\quad \forall\; P\in \Lambda^{\ssty A|BC}\cup\Lambda^{\ssty B|CA}\cup\Lambda^{\ssty C|AB}
	\label{product_equations_4}
\end{equation}

Using Gröbner basis formalism, we get the simplified form:
\begin{equation}
x_{\ssty5}=x_{\ssty6}=x_{\ssty7}=0
\end{equation}
along with the Eqs. (\ref{x_eqs}). This completes the proof. For brevity, we omit the corresponding lengthy calculations.

\section{XI. P\lowercase{roof of} P\lowercase{roposition} 5}

\textbf{\textit{Proposition:}}
The subspace spanned by  $\mathcal{U}$ forms an indistinguishable subspace.

Proof:
Let $\ket{\eta\down{1}{\ssty k}}$ and $\ket{\eta\down{1}{\ssty p}}$ be two distinct product states of the form given in (\ref{k_product}), corresponding to the $k$-th and $p$-th roots of Eq. (\ref{x_3_eq}). Then,
\begin{equation}
\zoom{c}{
	\begin{array}{lll}
	\braket{\eta\down{1}{\ssty k}}{\eta\down{1}{\ssty p}}&=\overline{x\down{0}{\ssty0}\up{0}{\ssty(k)}}x\down{0}{\ssty0}\up{0}{\ssty(p)}\braket{\phi\down{0}{\ssty 0}^{\ssty-}}{\phi\down{0}{\ssty 0}^{\ssty-}}+\overline{x\down{0}{\ssty1}\up{0}{\ssty(k)}}x\down{0}{\ssty1}\up{0}{\ssty(p)}\braket{\phi\down{0}{\ssty 1}^{\ssty-}}{\phi\down{0}{\ssty 1}^{\ssty-}}+\overline{x\down{0}{\ssty2}\up{0}{\ssty(k)}}x\down{0}{\ssty2}\up{0}{\ssty(p)}\braket{\psi\down{1}{\ssty 0}^{\ssty+}}{\psi\down{1}{\ssty 0}^{\ssty+}}\\[3pt]
	&\move{105}+\overline{x\down{0}{\ssty3}\up{0}{\ssty(k)}}x\down{0}{\ssty3}\up{0}{\ssty(p)}\braket{\psi\down{1}{\ssty 0}^{\ssty+}}{\psi\down{1}{\ssty 1}^{\ssty+}}+\braket{\kappa\,}{\kappa\,}\\[8pt]
	&=8+6\overline{x\down{0}{\ssty0}\up{0}{\ssty(k)}}x\down{0}{\ssty0}\up{0}{\ssty(p)}+14\overline{x\down{0}{\ssty1}\up{0}{\ssty(k)}}x\down{0}{\ssty1}\up{0}{\ssty(p)}+2\overline{x\down{0}{\ssty2}\up{0}{\ssty(k)}}x\down{0}{\ssty2}\up{0}{\ssty(p)}+2\overline{x\down{0}{\ssty3}\up{0}{\ssty(k)}}x\down{0}{\ssty3}\up{0}{\ssty(p)}\\[8pt]
	&\neq0,\quad\forall\;k\neq p\;(\text{Checked Numericaly})
	\end{array}
	}
\end{equation}

Therefore, no two states of the form (\ref{k_product}) are mutually orthogonal. The remaining arguments follow as discussed in the main text.
\end{document}